\documentclass[pra,aps,twocolumn,showpacs,amsmath,amssymb,superscriptaddress]{revtex4}
\usepackage[dvips]{color}
\usepackage{graphicx} 
\usepackage{bm}

\newcommand{\tr}{\mathop{\mathrm{Tr}}}
\newcommand{\im}{\mathop{\mathrm{Im}}}

\renewcommand{\d}{\mathrm{d}}

\newcommand{\be}{\begin{equation}}
\newcommand{\ee}{\end{equation}}
\newcommand{\bea}{\begin{eqnarray}}
\newcommand{\eea}{\end{eqnarray}}
\newcommand{\bse}{\begin{subequations}}
\newcommand{\ese}{\end{subequations}}
\renewcommand{\a}{a_{\mathrm{1D}}}
\newcommand{\chio}{\chi^{(0)}}
\newcommand{\pf}{k_{\mathrm F}}
\newcommand{\kf}{k_{\mathrm F}}
\newcommand{\ef}{\varepsilon_{\mathrm F}}
\newcommand{\gb}{g_\mathrm{B}}
\newcommand{\gf}{g_\mathrm{F}}
\newcommand{\q}{q}

\begin{document}
\title{Polarizability and dynamic structure factor of the one-dimensional Bose gas\\
near the Tonks-Girardeau limit at finite temperatures}

\author{Alexander Yu.~Cherny}
\affiliation{Max Planck Institute for the Physics of Complex
Systems, N\"othnitzer Stra{\ss}e 38, 01187 Dresden, Germany}
\affiliation{Bogoliubov Laboratory of Theoretical Physics, Joint
Institute for Nuclear Research, 141980, Dubna, Moscow region, Russia}

\author{Joachim Brand}
\affiliation{Max Planck Institute for the Physics of Complex
Systems, N\"othnitzer Stra{\ss}e 38, 01187 Dresden, Germany}

\date{January 12, 2006}

\begin{abstract}
Correlation functions related to the dynamic density response of the
one-dimensional Bose gas in the model of Lieb and Liniger are
calculated. An exact Bose-Fermi mapping is used to work in a fermionic
representation with a pseudopotential Hamiltonian. The Hartree-Fock
and generalized random phase approximations are derived and the
dynamic polarizability is calculated. The results are valid to first
order in $1/\gamma$ where $\gamma$ is Lieb-Liniger coupling
parameter. Approximations for the dynamic and static structure factor
at finite temperature are presented. The results preclude
superfluidity at any finite temperature in the large-$\gamma$ regime due
to the Landau criterion. Due to the exact Bose-Fermi duality, the
results apply for spinless fermions with weak $p$-wave interactions as
well as for strongly interacting bosons.
\end{abstract}


\maketitle

\section{Introduction}

The one-dimensional (1D) Bose gas with point interactions
is an abstract and simple, yet
nontrivial model of an interacting many-body system. The recent
progress in the trapping, cooling, and manipulation of atoms have made
it possible to test theoretical predictions experimentally and have
thus led to a revival of interest in this model.

Exact solutions
were described by Lieb
and Liniger~\cite{lieb63:1,lieb63:2}, who found that the physics
of the homogeneous Bose gas is
governed by the single dimensionless parameter $\gamma \equiv \gb m
/(\hslash^2 n)$. Here $n$ is the linear particle density, $m$ is the
mass, and $\gb$ is the strength of the short-range interaction
between particles~\cite{note:coupling}.
For small $\gamma$ we have a gas of weakly interacting bosons. Although
there is no Bose-condensation in one dimension even at zero temperature~\cite{hohenberg67},
many properties of the gas are reminiscent of
Bose-Einstein condensates with Bogoliubov perturbation theory being
valid and even superfluid properties were
predicted~\cite{sonin71,kagan00,buchler:100403}. For large $\gamma$, however,
the system crosses over into a strongly interacting regime and at
infinite $\gamma$ we obtain the Tonks-Girardeau (TG) gas of
impenetrable bosons \cite{girardeau60}. In this regime, the
strongly repulsive short-range interaction has the same effect as the
Pauli principle for fermions. Many properties like the excitation
spectrum become that of a free Fermi gas and, indeed, the model maps
one-to-one to a gas of noninteracting spinless fermions.

Experiments have
recently probed the crossover to the strongly-correlated TG
regime by increasing interactions up to values of $\gamma\approx 5.5$
\cite{Kinoshita2004} and to effective values of $\gamma_{\text{eff}}\approx
200$ in an optical lattice \cite{paredes04}. The momentum distribution
\cite{paredes04}, density profiles \cite{Kinoshita2004}, and low-energy
compressional excitation modes \cite{moritz:250402} have been the focus of the
experimental studies. In a recent experiment \cite{stoferle:130403}, the zero
momentum excitations of a 1D Bose gas in an optical lattice have been measured
by Bragg scattering, a technique that could also be used to measure the dynamic
structure factor (DSF) $S(q,\omega)$, which is calculated in this
paper~\cite{note_bragg}.

The theoretical description of the Lieb-Liniger model is not complete.
Although the exact wavefunctions,
the  excitation spectrum,
and the thermodynamic properties~\cite{yang69} are known for
arbitrary values
of the coupling constant $\gamma$, it is notoriously difficult to calculate the
correlation functions.
Many results in limiting cases are summarized in the
book~\cite{korepin93}, but the full problem is not yet solved. Recently,
progress has been made on the large-distance and long-time asymptotics
of single-particle correlation functions \cite{korepin97,essler98}
and on time-independent correlation functions
\cite{Gangardt2003a,Gangardt2003b,Shlyapnikov2003,olshanii02ep,astrakh03,muramatsu04,cheianov05}. In the
general case and for time-dependent correlation functions, a wealth of
information is available for small $\gamma$ where Bogoliubov perturbation
theory can be applied as well as for the TG gas at  $\gamma= \infty$. However,
the strongly-interacting
regime with large but finite $\gamma$ was hardly accessible as a systematic
expansion in $\gamma^{-1}$ was lacking.

In 1D there is a duality between interacting Bose and Fermi many-body
systems. A couple of recent works
\cite{cheon99,cazalilla03,granger03ep,girardeau03ep,grosse04,blume04}
already pointed out that the exact
Bose-Fermi mapping that Girardeau used to solve the case of
$\gamma=\infty$ \cite{girardeau60}
can be extended to the case of finite
interaction $\gamma$.
Thus, a  model system of interacting fermions can be constructed
for which the energy spectrum and associated wavefunctions are
in a one-to-one correspondence with
the Lieb-Liniger solutions. Our approach makes use of the same
Bose-Fermi duality, however, the motivation is to derive new results
for the 1D Bose gas.
We can calculate correlation functions of the strongly-interacting
Bose gas by solving the
equivalent interacting Fermi problem in the regime where its interactions are
small. In our previous paper~\cite{brand05} we calculated the DSF for
the Lieb-Liniger model for large $\gamma$ at zero temperature and
related it to the Landau criterion of superfluidity. The DSF
$S(q,\omega)$ holds information about the strength or the excitability
of excitations with momentum $\hslash q$ and energy $\hslash \omega$ and
thus may indicate decay routes of possible supercurrents. Although the
TG gas is not superfluid due to low-energy umklapp excitations near
$\omega=0$ and $q=2\pi n$, a crossover to superfluid behaviour for
finite $\gamma$ is possible if the umklapp excitations are suppressed.
This possibility is not precluded by the results of Ref.~\cite{brand05}.

The purpose of this paper is to calculate so far unknown
correlation properties of the 1D Bose gas at finite temperatures in
the strongly-interacting regime.
We calculate the dynamic
density-density response, the DSF, and the static structure factor of
the Lieb-Liniger gas with the fermionic random-phase approximation
(RPA), extending our previous results~\cite{brand05} to the case of
finite temperatures. Although our calculations are non-perturbative,
we obtain the first order term in the expansion in $\gamma^{-1}$ for
comparison. In particular we find that the 1D Bose gas at finite
temperature cannot be superfluid due to the finite probability of
umklapp excitations.
We also present an extended discussion of the validity of
the pseudopotential approach and give a critical scrutiny of the
limits of applicability of our approach.

The structure of this paper is as follows. In the next section we consider the
exact Bose-Fermi mapping for finite values of the interaction
strength and discuss the use of a fermionic pseudopotential.
In Sec.~\ref{sec:hf} we derive a
Hartree-Fock (HF) approximation for the fermions.
The generalized Random-Phase Approximation (RPA) is
derived in Sec.~\ref{sec:rpa} as a linearized time-dependent HF scheme.
Analytic expressions for the polarizability, the DSF, and the static
structure factor are analyzed and discussed.

\section{Fermionic pseudopotential}
\label{sec:mapping}

We consider the system of $N$ interacting bosons of mass $m$ in 1D
described by the Hamiltonian
\begin{align} \label{llham}
\hat{H}_{\rm B} =\sum_{i=1}^N \left[ -\frac{\hslash^2}{2 m}
\frac{\partial^{2}}{\partial x^{2}_i}
+ V_{\text{ext}}(x_i) \right]
      + \gb \sum_{i < j}
\delta(x_i - x_j).
\end{align}
This model extends the Lieb-Liniger model \cite{lieb63:1,lieb63:2} to
include an external potential  $V_{\rm ext}(x)$.
In contrast to the Bethe ansatz solutions of
Refs.~\cite{lieb63:1,lieb63:2}, our approach developed below allows
us to treat the effects of external potentials, which are important in
the context of experimental realizations.

We will now map the model (\ref{llham}) onto an equivalent Fermi
system and discuss the appropriate pseudopotentials. As described in
detail in Ref.~\cite{lieb63:1}, the $\delta$-function point
interaction in the Hamiltonian (\ref{llham}) can be represented as a
boundary condition for the wavefunction in coordinate space at the
points where two particles meet at the same position. For simplicity  we
will first discuss the case of two particles and introduce
center-of-mass $R=(x_{1}+x_{2})/2$ and relative $x=x_2-x_1$
coordinates. The bosonic wavefunction has the (even) symmetry
$\psi^{\rm B}(x,R) = \psi^{\rm B}(-x,R)$.
Due to this symmetry, the effect of the $\delta$-interaction on
$\psi^{\rm B}(x,R)$ can be formulated as a single boundary condition for $x\to
+0$:
\begin{align}
\lim_{x\to+0}
\partial_{x}\psi(x,R)
=\lim_{x\to+0}\frac{\gb m}{2\hslash^{2}}\psi(x,R) ,
\label{eqn:bbc}
\end{align}
with $\psi(x,R) = \psi^{\rm B}(x,R)$.
The exact Bose-Fermi mapping now takes advantage of this boundary condition
being formulated for $x>0$ where $\psi^{\rm B}(x,R)$ solves the
Schr\"odinger equation. A fermionic model is defined by the same
boundary condition (\ref{eqn:bbc}) and the same Schr\"odinger equation
for $x\neq 0$ but requiring fermionic (odd) symmetry. We thus find
fermionic solutions $\psi^{\rm F}$ with
\be
\psi^{\rm F}(x,R)=\left\{\begin{array}{ll}
\phantom{-}\psi^{\rm B}(x,R), &x>0,\\
-\psi^{\rm B}(x,R),           &x<0.
\end{array}\right.
\label{mapping}
\ee
The fermionic symmetry together with the boundary
conditions (\ref{eqn:bbc}) requires a discontinuity in the
wavefunction $\psi^{\rm F}(x,R)$ at $x=0$ inducing a jump of
$4\hslash^2/(\gb m) \partial_x\psi^{\rm F}(x=0,R)$ in the wavefunction
and a continuous first
derivative, whereas the bosonic
wavefunction $\psi^{\rm B}(x,R)$ is continuous but has a discontinuous
first derivative. In the simple limiting case of $\gb \to \infty$ we
obtain the TG gas and $\psi^{\rm F}$ is continuous.

This generalized Bose-Fermi mapping was introduced by Cheon and
Shigehara \cite{cheon99}, who also discussed the straightforward
generalization to the $N$-particle problem of Eq.~(\ref{llham}).
The mapping as described above is exact and one-to-one for particles
confined by an external potential
$V_{\rm ext}$. If periodic boundary conditions are imposed upon the
Bose system, they translate in the Fermi-system into
periodic boundary for odd $N$ and antiperiodic boundary for even $N$
[see Eq.~(37) of Ref.~\cite{cheon99}]. Antiperiodic boundaries mean
that the fermionic wavefunction changes by a factor of $-1$ whenever a
particle is translated by the length of the periodic box $L$. The
differences between periodic and antiperiodic boundaries are, however,
minute for large systems and vanish in the thermodynamic limit. For
this reason we only consider explicitly the case of periodic boundary
conditions for the fermionic wavefunctions below.

As a result of the Bose-Fermi mapping, the energy spectrum of the Bose
and corresponding Fermi system are
identical. Furthermore, all observables that are functions of the local density
operators are identical in both systems because they involve absolute
values of the
wavefunctions only and sign changes as in Eq.~(\ref{mapping}) do not matter. In
particular, this includes the dynamical density-density correlation functions
and derived quantities like the dynamic and static structure factors. By
contrast, the off-diagonal parts of the one-body density matrix and
consequently the momentum distribution show distinct differences in both
systems~\cite{lenard64}.

For our purposes it is desirable to represent the interaction in the fermionic
model as an operator. In fact, it has been shown by {\v S}eba that the
discontinuity-introducing boundary condition (\ref{eqn:bbc}) for fermionic
symmetry defines a self-adjoint operator on Hilbert space \cite{seba86}.  {\v
S}eba also gave an explicit construction by the zero-range limit of a
renormalized separable operator of finite range.  As a result, one can
represent the fermionic interaction in terms of an integral kernel
\cite{girardeau03ep,brand05,seba86}:
\begin{align}
\nonumber
&V_{\rm F}(x_1,x_2;x'_2,x'_1)= \\
& -{2}\gf \delta\left(
\frac{x_1+x_2-x'_1-x'_2}{2} \right)
\delta'(x_1-x_2)\delta'(x'_1-x'_2),
\label{eqn:our}
\end{align}
where the coupling constant in the fermionic representation is defined as
\be
\gf = {2 \hslash^4}/({m^2 \gb}).
\label{gf}
\ee
Due to the gap in the fermionic wave functions (\ref{mapping}), we should be
very careful defining the matrix element of the pseudopotential (\ref{eqn:our})
for two arbitrary fermionic two-particle wavefunctions:
\begin{align}
\langle\psi^{\rm F}|\hat{V}_{\rm F}|\varphi^{\rm F}\rangle
\!=\!-{2}\gf\!\int\d R\,&[\partial_{x}\psi^{{\rm F}}(x,R)]^{*}
 \partial_{x}\varphi^{\rm F}(x,R)\Big|_{x\to\pm 0} ,
\label{eqn:mat_el}
\end{align}
where due to the fermionic symmetry the right and left limits of the first
derivatives $\lim_{x\to\pm 0} \partial_{x}\psi^{{\rm F}}(x,R)$ coincide, even
if the wavefunctions are discontinuous at $x=0$. Representations similar to
Eqs.~(\ref{eqn:our}) and (\ref{eqn:mat_el}) have also been given in
Refs.~\cite{girardeau03ep,grosse04,blume04}. The fermionic Hamiltonian
$\hat{H}_{\rm F}$ takes the form of Eq.~(\ref{llham}) but with the interaction
term $\sum_{i<j}V_{\rm F}(x_i,x_j;x'_j,x'_i)$ instead of the bosonic
$\delta$-function interactions. If the bosonic wavefunction $\psi^{\rm B}_{n}$
is an eigenfunction of the Hamiltonian $\hat{H}_{\rm B}$ of Eq. (\ref{llham})
with eigenvalue $E^{\rm B}_{n}$, then the fermionic wavefunction $\psi^{\rm
F}_{n}$ of Eq.~(\ref{mapping}) is also eigenfunction of the fermionic
Hamiltonian with interaction $V_{\rm F}$ of Eq.~(\ref{eqn:our}) corresponding
to the same eigenvalue $E^{\rm F}_{n} =E^{\rm B}_{n}$. This can be verified
easily for two particles by substituting $\psi^{\rm F}_{n}$ into the
Schr\"odinger equation. We thus conclude that the fermionic representation with
interactions (\ref{eqn:our}) is exact for all values of the coupling constant
$\gf$.

In addition to the formal considerations above we now give another, somewhat
heuristic  justification for the pseudopotential
(\ref{eqn:our}) following Sen's argument \cite{sen99} based on the
Hellmann-Feynman theorem. Let us suppose that we know the exact eigenvalue
$E_{n}^{{\rm B}}$ of $\hat{H}_{\rm B}$ together with the corresponding bosonic
wave function $\psi^{\rm B}_{n}$. Then it follows from the Hellmann-Feynman
theorem that
\begin{align}
\frac{\partial E_{n}^{{\rm B}}}{\partial\gb}
=&\ \bigg\langle\psi^{\rm B}\bigg|\frac{\partial\hat{H}_{\rm B}}{\partial\gb}\bigg|
\varphi^{\rm B}\bigg\rangle =\int\d R\,|\psi^{\rm B}(x=\pm 0,R)|^{2}\nonumber\\
=&\ \int\d R\,\frac{4\hslash^{4}}{m^{2}\gb^{2}}|\partial_{x}\psi^{\rm F}(x=\pm 0,R)|^{2},
\label{eqn:den}
\end{align}
where Eqs.~(\ref{eqn:bbc}) and (\ref{mapping}) were used to derive the last
equality. With the help of Eqs.~(\ref{gf}) and (\ref{eqn:mat_el}) we find
${\partial E_{n}^{\rm B}}/{\partial\gf}=-2\langle\psi^{\rm
F}|{\partial\hat{V}_{\rm F}}/ {\partial\gf}|\psi^{\rm F}\rangle$ and finally
${\partial E_{n}^{\rm B}}/{\partial\gf}={\partial E_{n}^{\rm F}}/{\partial\gf}$
by the  Hellmann-Feynman theorem for the eigenvalue $E_{n}^{\rm F}$ of
$\hat{H}_{\rm F}$. Taking into account that $\hat{H}_{\rm B}$ and $\hat{H}_{\rm
F}$ have the same eigenvalue spectrum in the TG limit of  $\gf\to0$, we
conclude that both Hamiltonians have the same spectrum also for arbitrary
values of $\gf$.

A useful {\it approximate} representation as a pairwise
pseudopotential was suggested by Sen~\cite{sen99}:
\be \label{eqn:SenPP}
  V_{\rm Sen}(x_1,x_2) =
  -\gf \delta''(x_1-x_2),
\ee
where $\delta''(x)$ denotes the second derivative of the delta function. In
contrast to the integral kernel (\ref{eqn:our}), Sen's pseudopotential takes
the form of a {\it local} operator, which yields a simplification when
performing analytical calculations. This pseudopotential, however, is
applicable only for variational calculations in a variational space of
continuous fermionic functions that vanish whenever two particle coordinates
coincide.  This is the case for Slater determinants that may be used to derive
the Hartree-Fock (HF) and Random-Phase approximations (RPA) but not for the
exact fermionic wave functions like (\ref{mapping}). One can justify Sen's
pseudopotential (\ref{eqn:SenPP}) up to first order in $\gamma^{-1}$ in the
same manner as in the previous paragraph. For higher orders, $V_{\rm Sen}$ is
not correct, because its matrix elements contain not only the correct term, as
in the r.h.s.\ of Eq.~(\ref{eqn:mat_el}), but an additional term disappearing
only at $\gamma^{-1}=0$.

The fermionic Hamiltonian can now be rewritten
in terms of Fermi field operators ${\hat\Psi}(x)$ and ${\hat\Psi}^{\dag}(x)$
\begin{align}
\hat{H}_{\rm F}=&\int \d
x\frac{\partial_{x}{\hat\Psi}^{\dag}(x)\partial_{x}{\hat\Psi}(x)}{2m}
+\int \d x\,
V_{\rm ext}(x){\hat\Psi}^{\dagger}(x){\hat\Psi}(x)\nonumber\\
&+\frac{1}{2}\int \d x_1 \d x_2 \d x'_1 \d x'_2\,
V_{\rm F}(x_1,x_2;x'_2,x'_1)\nonumber\\
&\times{\hat\Psi}^{\dagger}(x_{1}){\hat\Psi}^{\dagger}(x_{2})
{\hat\Psi}(x'_{2}){\hat\Psi}(x'_{1}),
\label{hamfermi}
\end{align}
where $V_{\rm F}$ is given by Eq.~(\ref{eqn:our}). Alternatively, the
approximate pseudopotential $V_{\rm Sen}$ can be employed~\cite{note_local}.

In the remainder of this
paper we will study the fermionic model (\ref{hamfermi}) in the HF
approximation and the RPA.

\section{The Hartree-Fock operator}
\label{sec:hf}

The HF approximation for the fermionic system (\ref{hamfermi}) is
derived in the standard way by variation over Slater determinants. We
thus expect Sen's pseudopotential  (\ref{eqn:SenPP}) to be
valid. Indeed, we find that the interactions  (\ref{eqn:SenPP}) and
(\ref{eqn:our}) yield identical results on the HF level.

When working at finite temperatures, it is convenient to introduce the
HF operator as the single-particle operator
\begin{equation}
\hat{H}_0=\int \d x\,\d x'\,F(x,x'){\hat\Psi}^\dag(x){\hat\Psi}(x')
\label{hf}
\end{equation}
that minimizes the Gibbs-Bogoliubov inequality \cite{isihara68}
with respect to $F(x,x')$
$$
\Omega\leqslant \Omega_0 +\langle\hat{H}_{\rm F}-\hat{H}_0\rangle_0.
$$
Here $\Omega=-\frac{1}{\beta}\ln Z$ is the grand thermodynamic
potential with the partition function $Z\equiv\tr
\exp[-\beta(\hat{H}_{\rm F}-\mu\hat{N})]$ corresponding to the Hamiltonian
$\hat{H}_{\rm F}$. The inverse temperature $\beta=1/T$ is introduced here and we
use the units $k_B=1$ in this paper. Accordingly, $\Omega_0$ is associated with $\hat{H}_{0}$,
and $\langle\cdots\rangle_0\equiv\frac{1}{Z_0}\tr\{
\cdots\exp[-\beta(\hat{H}_0-\mu\hat{N})]\}$. The variational procedure yields
\begin{align}
&F(y,z)= \delta(y-z)\left[-\frac{\hslash^2}{2m}\frac{\partial^{2}}{\partial z^{2}}+V_{\rm ext}(z,t)\right]\nonumber\\
&+\int \d x\d x'\,\left[V(y,x;x',z)-V(y,x;z,x')\right] {\rho}^{(1)}(x,x'),
\label{tconv1}
\end{align}
with the one-body density matrix ${\rho}^{(1)}(x',x)
\equiv\langle{\hat\Psi}^\dag(x') {\hat\Psi}(x)\rangle_0$, which should be
determined in a self-consistent manner.  The simplest way to do this
is to work in the diagonal representation of
the HF kernel
$F(x,x')=\sum_{j}\varepsilon_{j}\varphi^{*}_{j}(x')\varphi_{j}(x)$.
The single-particle functions $\varphi_{j}(x)$ are called Hartree-Fock
orbitals. This representation allows us to rewrite
the HF Hamiltonian (\ref{hf}) in terms of the creation and destruction operators
$\hat{a}^{\dag}_{j}\equiv\int \d x\,\hat{\Psi}^{\dag}(x)\varphi_{j}(x)$ and
$\hat{a}_{j}\equiv\int \d x\,\hat{\Psi}(x)\varphi^{*}_{j}(x)$, respectively. It takes
the form $\hat{H}_0=\sum_{j}\varepsilon_{j}\hat{a}^{\dag}_{j}\hat{a}_{j}$, which
leads to
\be
\langle\hat{a}^{\dag}_{i}\hat{a}_{j}\rangle_{0}=n_j\delta_{ij},
\label{ijcorr}
\ee
where
\be
n_j=\frac{1}{\exp[\beta(\varepsilon_j-\mu)]+1}.
\label{oc_num}
\ee
is the Fermi distribution of occupation numbers, and $\delta_{ij}$ is the
Kroneker symbol. At zero temperature, $n_j$ defines the Fermi step
function. By using
the representation $\hat{\Psi}(x)=\sum_{j}\hat{a}_{j}\varphi_{j}(x)$ and
Eq.~(\ref{ijcorr}), we derive
\be
{\rho}^{(1)}(x,x')
=\sum_j n_j\varphi^{*}_{j}(x)\varphi_{j}(x').
\label{rho1_exp}
\ee

By substituting Eq.~(\ref{rho1_exp}) and either one of the the
pseudopotentials (\ref{eqn:SenPP}) or
(\ref{eqn:our}) into Eq.~(\ref{tconv1}), we come to the same {\it
local} form of the HF kernel
\begin{align} \label{eqn:HFoperator}
\hat{F}=&-\frac{\hslash^2}{2m}\frac{\partial^2}{\partial x^2}+V_{\rm
ext}(x)\nonumber\\
&+
\gf \left[n(x)\frac{\partial^2}{\partial x^2}+
{2}{\cal P}(x)i\frac{\partial}{\partial x}
-{\cal T}(x)\right],
\end{align}
which is defined as  $\hat{F}\varphi(x)\equiv\int \d x'\,F(x,x')\varphi(x')$.
The first two terms on the right hand side come from
the single-particle part of the Hamiltonian~(\ref{hamfermi}).
The mean-field parts in the square bracket involve the
local density $n(x) = {\rho}^{(1)}(x,x)=\sum_j n_j
{{\varphi}^{*}}_{j}(x)\varphi_{j}(x)$ and the derivative densities ${\cal
P}(x)\equiv-i\sum_j n_j {{\varphi}^{*}}'_{j}(x)\varphi_{j}(x)$ and ${\cal
T}(x)\equiv\sum_j n_j [{\varphi}^{*}_{j}(x)\varphi''_{j}(x) +
2{{\varphi}^{*}}'_{j}(x)\varphi'_{j}(x)]$. Here we define $\varphi' \equiv \d
\varphi/\d x$. The quantities ${\cal P}(x)$ and ${\cal T}(x)$ are
reminiscent of momentum and energy densities, respectively.   We find a purely
local Fock operator, contrary to the case of Coulomb interactions where the
Fock operator $\hat{F}$ is nonlocal with a local Hartree and a nonlocal
exchange term.

For the homogeneous gas ($V_{\rm ext}= 0$), the quantum number $j$ can be
associated with the particle
HF orbitals are plane waves
$\varphi_{q}(x)=\exp(ixq)/\sqrt{L}$ with energy and effective mass
\begin{align}
\varepsilon_q=\ &\frac{\hslash^2 q^2}{2m^*}-\gf n\langle q^{2}\rangle,
\label{eqn:epsHF}\\
m^* \equiv\ & \frac{m}{1-{2 \gf m n}/{\hslash^2}} = \frac{m}{1-4
   \gamma^{-1}},
\label{meff}
\end{align}
respectively, where we have introduce the average square momentum
\be
\langle q^{2}\rangle\equiv\frac{1}{N}\sum_{q}n_{q} q^{2}
\label{q2av}
\ee
over the Fermi distribution $n_q$ of Eq.~(\ref{oc_num}).
Here, the sum over $q$ runs over values $q=2\pi l/L$, $l=0,\pm
1,\pm 2,\cdots$ in accordance with  periodic boundary conditions.
The chemical potential and the density cannot be independent quantities;
they are related through
\be
n=\frac{1}{L}\sum_{q} n_{q}.
\label{n_mu}
\ee
In the thermodynamic limit $n=N/L={\rm const}$, $L\to\infty$  all the sums over
momentum become integrals: $({1}/{L})\sum_{q} \to (2\pi)^{-1} \int \d q \cdots$.
In the canonical ensemble only two thermodynamic parameters are independent,
the  density and temperature. Thus, we can use $\gamma$ and $\beta$ as input
parameters and determine the chemical potential $\mu$ in a self-consistent
manner from Eqs.~(\ref{eqn:epsHF})-(\ref{n_mu}).

At zero temperature the HF scheme admits the analytical solutions $\langle
q^{2}\rangle=\kf^{2}/3$, $\mu=\varepsilon_{\kf}=\ef [1-16/(3\gamma)]$, where
$\kf=\pi n$ and  energy $\ef =  \hslash^2 \kf^2/(2 m)$ are Fermi
wavenumber and energy of the TG gas, respectively. The HF approximation for the
ground-state energy
yields
\be \label{eqn:EHF}
E_{\mathrm{HF}} \equiv \langle\hat{H}_{\rm F}\rangle_0=N \frac{\hslash^2\pi^2 n^2}{6 m} (1 - 8
\gamma^{-1}).
\ee
It coincides with the first two terms of the large-$\gamma$ expansion of the
exact ground-state energy in the Lieb-Liniger model \cite{lieb63:1}.
Note that  $E_{\rm
HF}\not=\langle\hat{H}_0\rangle_0=\sum_{p}\varepsilon_p$, as it should be in
the HF scheme (see e.g. Ref.~\cite{pines89}).

\begin{figure}
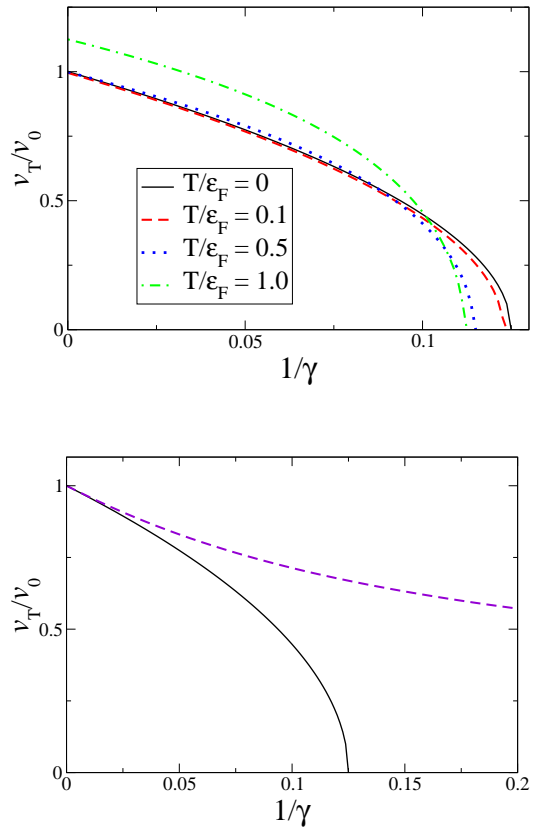

\includegraphics[width=0.8\columnwidth,clip=on]{vsa1.eps}\\[.9cm]
\includegraphics[width=0.8\columnwidth,clip=on]{vsb1.eps}
\caption{\label{fig:vsound}
(Color online) (a) The isothermal speed of sound $v_{\rm T}$ of Eq.~(\ref{eqn:speed}) versus
the inverse coupling
constant $\gamma^{-1}$ in the HF approximation for different
temperatures.
The quantity
$v_{0}\equiv\hslash \pi n /m$ is the speed of sound of the TG gas at zero
temperature.
The speed of sound becomes zero at some critical value of $\gamma$,
below which the HF solution becomes unstable (see discussion in
Sec.~\ref{sec:hf}.) (b) The solid (black) line shows the speed of
sound in the HF approximation at $T=0$ given by
Eq.~(\ref{eqn:speedLL}), and the dashed (violet) line shows the exact
speed of sound in the Lieb-Liniger model \cite{lieb63:2} for
comparison.}
\end{figure}

We now briefly discuss the stability of the HF solution.
Stability of the HF solution implies positivity of the isothermal
compressibility of the medium $(\partial n/\partial \mu)_{T}/n$. The latter
relates directly to the isothermal speed of sound
\be
v_{\rm T}=\sqrt{\frac{n}{m}\Big(\frac{\partial \mu}{\partial n}\Big)_{T}},
\label{eqn:speed}
\ee shown in Fig.~\ref{fig:vsound} for various temperatures. The HF
result at $T=0$
\begin{align}
\label{eqn:speedLL}
v_{\rm T} &= \frac{\hslash \pi n}{m} \sqrt{1- 8 \gamma^{-1}}\\
\nonumber &= \frac{\hslash \pi n}{m} \left[1- 4 \gamma^{-1} + {\cal
O}(\gamma^{-2})\right]
\end{align}
yields the correct first order expansion of Lieb's exact result
\cite{lieb63:2}.
We see in  Fig.~\ref{fig:vsound} that the stability condition
$v^{2}_{T} > 0$ is broken
below some critical value of the coupling constant $\gamma$, depending
on temperature. From Eqs.~(\ref{eqn:epsHF})-(\ref{n_mu}) one can show
that the critical value lies between $\gamma = 8$ at zero temperature and
$\gamma = 4\sqrt{3}/(1-\sqrt{3}) \approx 9.464$ at large
temperatures. Thus, the
developed HF scheme (and, hence, the RPA discussed below) is
applicable only for values of the Lieb-Liniger coupling constant of the order
$\gamma\gtrsim 10$.

\section{The random phase approximation}
\label{sec:rpa}

\subsection{Response function}
\label{sec:polar}

The HF approximation permits us to calculate the linear response of
time-dependent HF. Approximations of the linear response functions on
this level are known as RPA with exchange or generalized RPA
\cite{pines89,pines61}. In this section we will calculate the
density-density response function $\chi(q,z)$ also
known as dynamic polarizability. It is intimately related to the
DSF and the
time-dependent density-density correlation function
\cite{pines89,pines61,pitaevskii03:book}.  In order to define
$\chi(q,z)$ we consider
the linear response of the density
$$
{n}(x,t)-n=\langle{\hat\Psi}^\dag(x,t){\hat\Psi}(x,t)\rangle-{n}=\frac{1}{L}\sum_{q}e^{iqx}\delta
{n}(q,t)
$$
to an infinitesimal time-dependent external potential
$$
\delta V_{\mathrm{ext}}(x,t)=
\sum_{q}\int \hslash\frac{\d\omega}{2\pi}
e^{iqx}e^{-i\omega t}e^{\varepsilon t}\delta V_{\mathrm{ext}}(q,\omega).
$$
Here we choose $\varepsilon\to+0$ to provide the boundary condition
$V_{\mathrm{ext}}(x,t)\to 0$ when $t\to-\infty$. For $q\not=0$ we have
\begin{align}
\delta{n}(q,t)=&\ \sum_k\langle\hat{a}^{\dag}_{k-q/2}(t)\hat{a}_{k+q/2}(t)\rangle
\nonumber\\
              =&\ \int \hslash\frac{\d\omega}{2\pi}
e^{-i\omega t}e^{\varepsilon t}\delta{n}(q,\omega).
\label{denresp}
\end{align}
The dynamic polarizability
is now defined by
\begin{equation}
\chi(q,\omega+i\varepsilon)\equiv{-\delta n(q,\omega)}/{\delta V_{\mathrm{ext}}(q,\omega)}
\label{chiqom}
\end{equation}
and obviously determines the linear density response to an external
field.

The polarizability can be obtained  directly from the linearized
equation of motion of the density operator in the time-dependent HF
approximation in the standard way as summarized below.

({i})
With the help of the HF Hamiltonian (\ref{hf}) we write the equation of motion
$i\hslash\partial\hat{\rho}^{(1)}/\partial t=[\hat{\rho}^{(1)},\hat{H}_{0}]$
for the operator
$\hat{\rho}^{(1)}\equiv{\hat\Psi}^\dag(y,t){\hat\Psi}(z,t)$ and take its average.
We thus derive
\begin{align}
i\hslash\frac{\partial{\rho}^{(1)}(y,z,t)}{\partial t}=\int \d
x\,\big[&F(z,x){\rho}^{(1)}(y,x,t) \nonumber \\
&-F(x,y){\rho}^{(1)}(x,z,t)\big]
\label{denseq}
\end{align}
with the HF kernel $F$ of Eq.~(\ref{tconv1}).

({ii})
We substitute
${\rho}^{(1)}(y,z,t)
={\rho}^{(1)}_{0}(y-z)+\delta{{\rho}^{(1)}}(y,z,t)$ into Eq.~(\ref{denseq}) and linearize it
with respect to $\delta{\rho}^{(1)}$ and $\delta V_{\mathrm{ext}}$. Here we introduce the equilibrium value
of the one-body density matrix
${\rho}^{(1)}_{0}(y-z)=(1/L)\sum_{k}n_{k}\exp[ik(y-z)]$ in the HF approximation
with the HF occupation numbers $n_k$ of Eq.~(\ref{oc_num}).

({iii})
In the Fourier representation of momentum and frequency, the
obtained linearized equation becomes algebraic
and takes the form
\begin{align}
&\delta\tilde{\rho}^{(1)}(k,q,\omega)\bigg(
\frac{1}{L}\!\sum_{p}\!n_{p}\big[\mathcal{V}(k-p-q/2)-\mathcal{V}(k-p+q/2)\big]\nonumber\\
&\phantom{\delta\tilde{\rho}^{(1)}(k,q,\omega)\bigg(}
+\frac{\hslash^{2}kq}{m}-\hslash\omega- i\varepsilon\bigg) \nonumber \\
&=\bigg(\delta V_{\mathrm{ext}}(q,\omega)+\frac{1}{L}\sum_{p}\big[ \mathcal{V}(q)-\mathcal{V}(p-k)
\big]\delta\tilde{\rho}^{(1)}(p,q,\omega)\bigg)\nonumber\\
&\phantom{=}\ \times (n_{k+q/2}-n_{k-q/2}).
\label{fkq}
\end{align}
Here, $\mathcal{V}(q)=\gf q^2$ stands for the Fourier transform of
the potential (\ref{eqn:SenPP})  and $\delta\tilde{\rho}^{(1)}$ is defined by
the relation $\langle\hat{a}^{\dag}_{k-q/2}(t)\hat{a}_{k+q/2}(t)\rangle =(\hslash/{2\pi})\int
 \d\omega e^{-i\omega t} e^{\varepsilon t} \delta\tilde{\rho}^{(1)}(k,q,\omega)$
for $q\not=0$.
We are interested in the density response $\delta n(q,\omega)$,
which is
directly connected to $\delta\tilde{\rho}^{(1)}$ by
\begin{equation}
\delta
n(q,\omega)=\sum_{k}\delta\tilde{\rho}^{(1)}(k,q,\omega).
\label{nqom}
\end{equation}

Because $\mathcal{V}(q)$ is a
polynomial in $q$, we can obtain an analytical expression for the
polarizability (\ref{chiqom}) from Eqs.~(\ref{fkq}) and (\ref{nqom}). After
somewhat lengthy but straightforward calculations we find
\begin{equation}
\chi(q,z) = \frac{\chio(q,z)}
{(1-4\gamma^{-1}) [B+D(q,z)\chio(q,z)]}
\label{chqom1}
\end{equation}
with $z\equiv\omega+i\varepsilon$ and $\varepsilon {\to} +0$.
Here we denote
\begin{align}
B\equiv\ &1-{4\,(3\,\gamma -16)}/{(\gamma-4)^3} ,\nonumber \\
  D(q,z)\equiv\ &\frac{4 \ef}{N}\frac{\gamma}{(\gamma -4)^2}\Bigg[\frac{q^2}{\pf^2}
  \frac{2\gamma - 9} {2 \gamma}
  - \frac{6}{\gamma}\frac{\langle q^2\rangle}{\kf^2}
\nonumber \\
  & - \left(\frac{\hslash z \pf}{
  \ef q} \right)^2 \frac{3\gamma -16}{2( \gamma  -4)^2}\Bigg],\nonumber
\end{align}
and the polarizability $\chio$ of the ideal 1D
Fermi gas with renormalized mass is given by the relation
\be
\chio(q,z)=\sum_{k}
\frac{n_{k+q/2}-n_{k-q/2}}{\hslash z -{\hslash^{2}kq}/{m^{*}}}.
\label{chi_id}
\ee

\begin{figure}
\includegraphics[width=0.9\columnwidth]{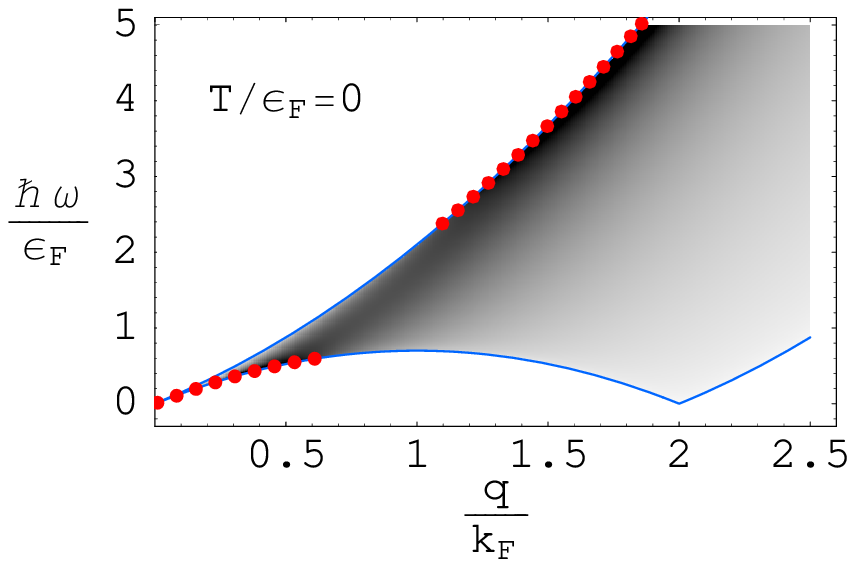}
\includegraphics[width=0.9\columnwidth]{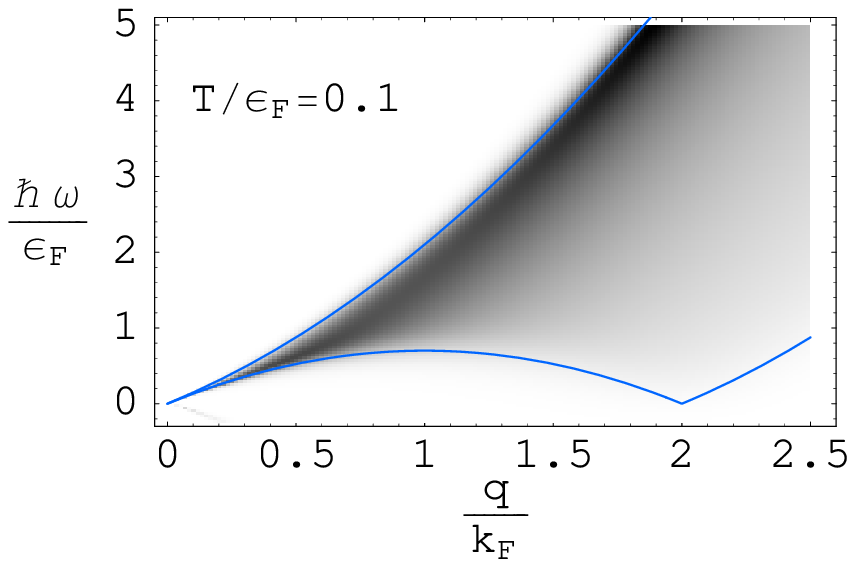}
\includegraphics[width=0.9\columnwidth]{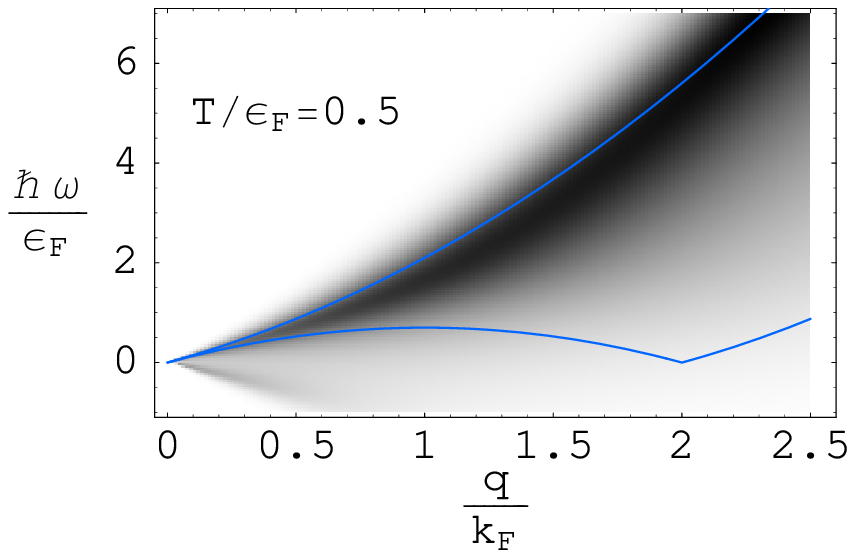}
\caption{\label{fig:3D-DSF}
(Color online) The excitation spectrum and the DSF at $\gamma = 13$ for various temperatures.
The upper and lower thin (blue) lines show the dispersions $\omega_+(q)$ and
$\omega_-(q)$ of Eq.~(\ref{eqn:dispersions}), respectively, limiting the
elementary excitations of the Lieb-Liniger model at $T=0$.
The dimensionless value of the rescaled DSF
$S(q,\omega)q\varepsilon_{F}/(\kf N)$ from Eqs.~(\ref{eqn:strdyn}) and
(\ref{eqn:strdyn_t}) is shown in shades of
grey between zero (white) and 1.0 (black). Here $\kf=\pi n$
and $\ef=\hslash^2 \kf^2 /(2 m)$. The dotted (red) line indicates a $\delta$-function
contribution at $\omega_0(q)$. At non-zero temperatures the
$\delta$-function contribution washes out and becomes a part of the continuum.}
\end{figure}

In  the thermodynamic limit we find the real and imaginary parts of
$\chio(q,\omega+i\varepsilon)=\chio_{1}(q,\omega)+i\chio_{2}(q,\omega)$
using the relation $1/(x+i\varepsilon)={\rm
P}({1}/{x})- i\pi\delta(x)$
\begin{align}
\chio_{1}(q,\omega)=&\ \frac{N m^{*}}{2\hslash^2 q \kf}\,{\rm P}\int\d k\,
\frac{n_{k+q_{-}}-n_{k+q_{+}}}{k},
\label{chi1t}\\
\chio_{2}(q,\omega)=&\ \frac{N m^{*}}{2\hslash^2 q \kf}\pi(n_{q_{-}}-n_{q_{+}}),
\label{chi2t}
\end{align}
where ${\rm P}$ means the Cauchy principal value and we defined
\be
q_{\pm}\equiv\frac{\omega m^{*}}{\hslash q}\pm \frac{q}{2}.
\label{kpm}
\ee
At zero temperature
the occupation numbers $n_{k}$ define the Fermi step function
and we arrive
at the simple analytic expressions
\begin{align}
  \chi^{(0)}_{1}(q,\omega)=\ &\frac{N m^{*}}{2\hslash^2 q \kf}\ln\left|
  \frac{\omega^{2}_{+}(q) - \omega^{2}}{\omega^{2}_{-}(q)-\omega^{2}}
  \right|, \label{chi1}\\
  \chi^{(0)}_{2}(q,\omega)=\ & \frac{N \pi m^{*}}{2\hslash^2 q \kf}\left\{
  \begin{array}{ll}
  \pm 1,& \omega_{-}\leqslant \pm \omega\leqslant \omega_{+},\\[2mm]
  0,& \text{otherwise.}
\end{array}\right.
\label{chi2}
\end{align}
The dispersion relations
\be
\omega_\pm(q)\equiv {\hslash |2 \pf q \pm q^2|}/{(2 m^*)}
\label{eqn:dispersions}
\ee
border the  continuum part of the accessible excitation spectrum made up from
HF quasiparticle-quasihole excitations~(\ref{eqn:epsHF}), as shown in
Fig.~\ref{fig:3D-DSF}. The two branches $\omega_\pm(q)$ thus
approximate the two branches of elementary excitations introduced by
Lieb \cite{lieb63:2} as type~I and type~II excitations, respectively.

In accordance with the exact
results, both branches share the same slope at the origin and give rise to a
single speed of sound  at zero temperature given by
$v_{\rm T} = {\d \omega_\pm}/{\d k} = {\hslash \kf}(1-4\gamma^{-1})/{m}$.
This value is
the correct first order expansion~\cite{lieb63:2} of $v_{\rm T}$ for large $\gamma$,
consistent with  Eq.~(\ref{eqn:speedLL}).
Note that the usual Bogoliubov perturbation theory \cite{bog47} for weakly interacting
bosons gives a similar expansion of $v_{\rm T}$ for small $\gamma$ and the type I
excitation branch. Type II excitations are not described with Bogoliubov
theory. The dispersion curves $\omega_\pm(q)$ of Eq.~(\ref{eqn:dispersions})
differ from the free Fermi gas (TG gas) values only by the renormalization
of the mass, which already takes place in the HF single-particle energies.

\subsection{Dynamic structure factor}
\label{sec:dsf}

The DSF
$S(q,\omega)$ is the Fourier transform of the density-density
correlation function~\cite{pines89,pines61,pitaevskii03:book} and expresses the
probability to excite a particular excited state through a density
perturbation
\begin{equation}
S(q,\omega)={\cal Z}^{-1}\sum_{n,m}e^{-\beta E_{m}}|\langle m|\hat{\rho}_{q}|n \rangle|^{2}
\delta(\hslash \omega-E_{n}+E_{m}) ,
\label{sqomega}
\end{equation}
where $\hat{\rho}_{q}=\sum_i \exp(-i q x_i)$ is the Fourier component of the
density operator, ${\cal Z}=\sum_{n}\exp(-\beta E_{n})$ is the
partition function.

The DSF is related to the dynamic polarizability by
\begin{equation}
\chi(q,\omega+i\varepsilon)
=\int_{-\infty}^{+\infty}\d\omega'\frac{2\omega'S(q,\omega')}{{\omega'}^{2}-(\omega+i\varepsilon)^{2}}
\label{chiqomandsqom}
\end{equation}
or, equivalently
\cite{pitaevskii03:book}, by
\begin{equation}
S(q,\omega)=\frac{\im\chi(q,\omega+i\varepsilon)}{\pi[1-\exp(-\beta\hslash\omega)]},
\label{imchisqomega}
\end{equation}
which gives at zero temperature
\begin{equation}
S(q,\omega)=\left\{\begin{array}{ll}
\im \chi(q,\omega+i\varepsilon)/\pi, &\omega>0,\\
0,                                    &\omega<0.

\end{array}\right.
\label{strsusc}
\end{equation}

\subsubsection{Zero temperature}

The DSF at zero temperature can be obtained from Eqs.~(\ref{chqom1})
and (\ref{strsusc}), which result in
\begin{align}
  S(q,\omega)=&\frac{ \chi^{(0)}_{2}(q,\omega) B} {\pi (1-4\,\gamma^{-1})
               \left[\Big(B+D\chi_{1}^{(0)}\Big)^2 +
               \Big(D\chi_{2}^{(0)}\Big)^2\right]}\nonumber\\
              &+\delta[\omega - \omega_0(q)] A(q)/\hslash,
\label{eqn:strdyn}
\end{align}
with $\chio_{1,2}$ given by the zero temperature expressions
(\ref{chi1}) and (\ref{chi2}).
A grey scale plot of this result is shown in Fig.~\ref{fig:3D-DSF}.
The DSF of Eq.~(\ref{eqn:strdyn}) has two contributions. The first part is
continuous and takes nonzero (and positive) values only for $\omega_-
< \omega <\omega_+$, which is also the region where particle-hole
excitations on the HF level are present. The second part is a discrete
branch with strength $A(q)$ and located at
$\omega = \omega_0(q)$, outside the region of the discrete
contribution. As we will discuss in detail below, the discrete part is
exponentially suppressed for small $\gamma^{-1}$ and should be
understood as an artefact of the RPA approximation.

\begin{figure*}
\noindent\includegraphics[width=.22\textwidth,angle=270,clip=on]{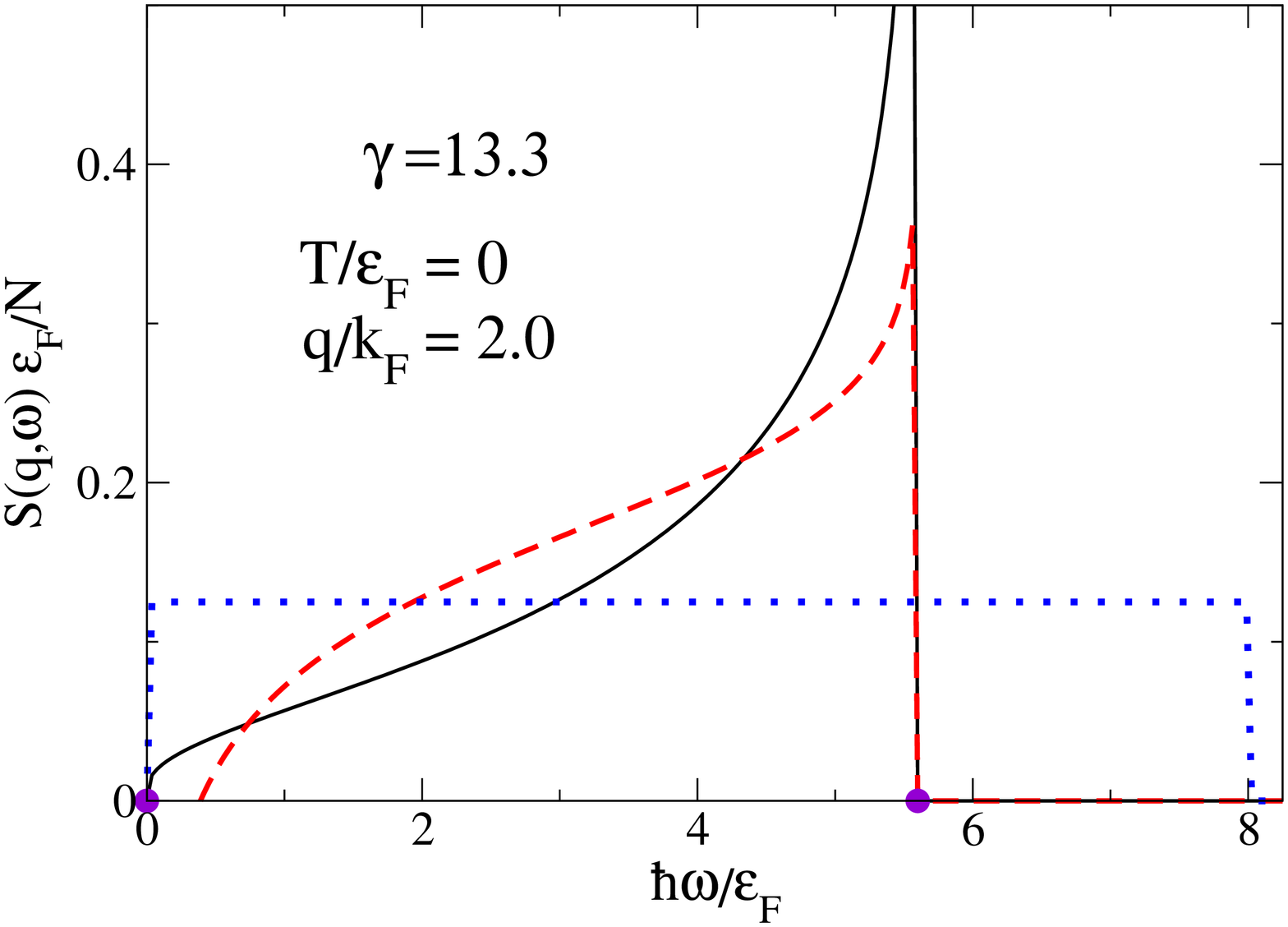}
\ \includegraphics[width=.22\textwidth,angle=270,clip=on]{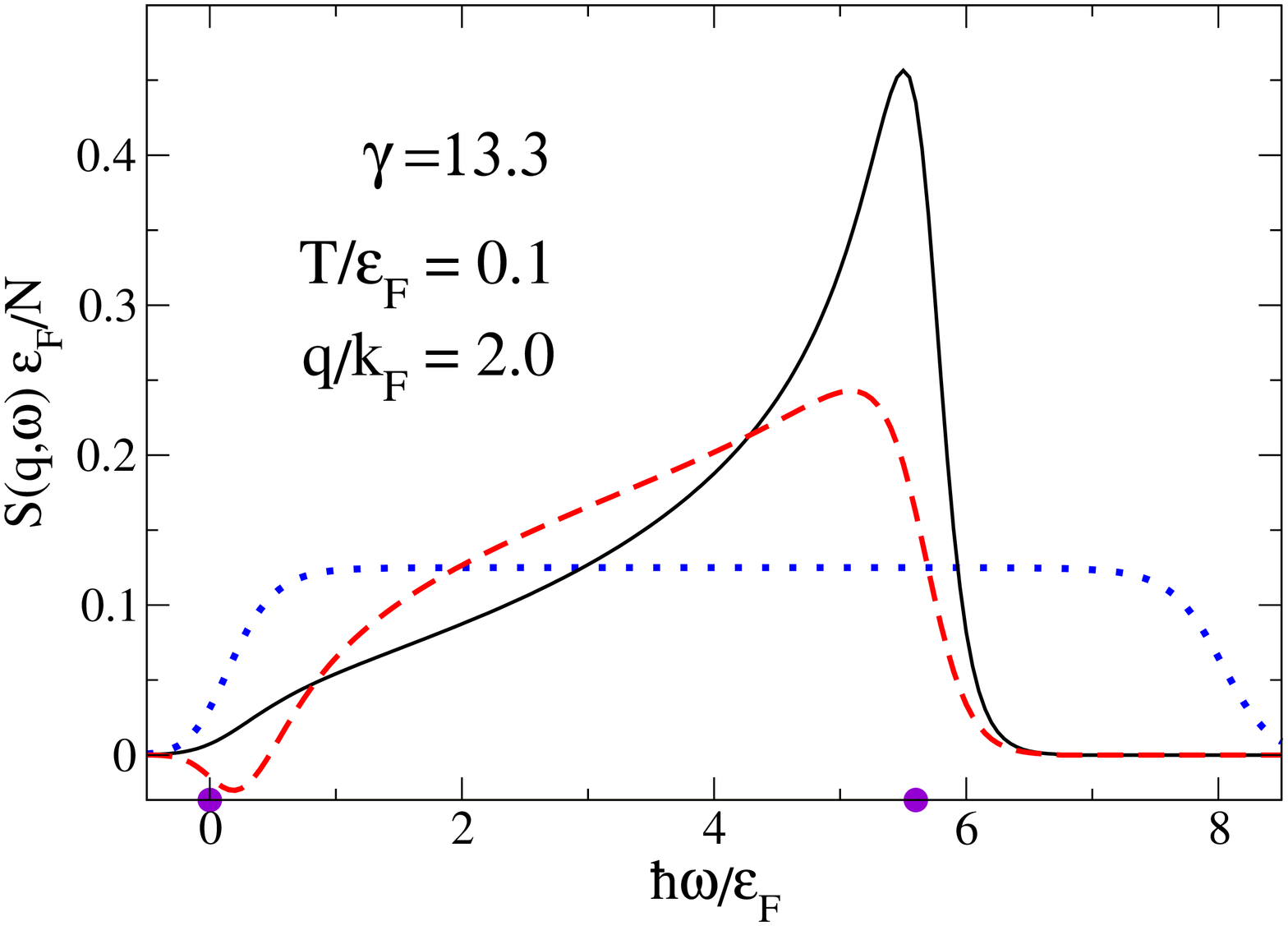}
\ \includegraphics[width=.22\textwidth,angle=270,clip=on]{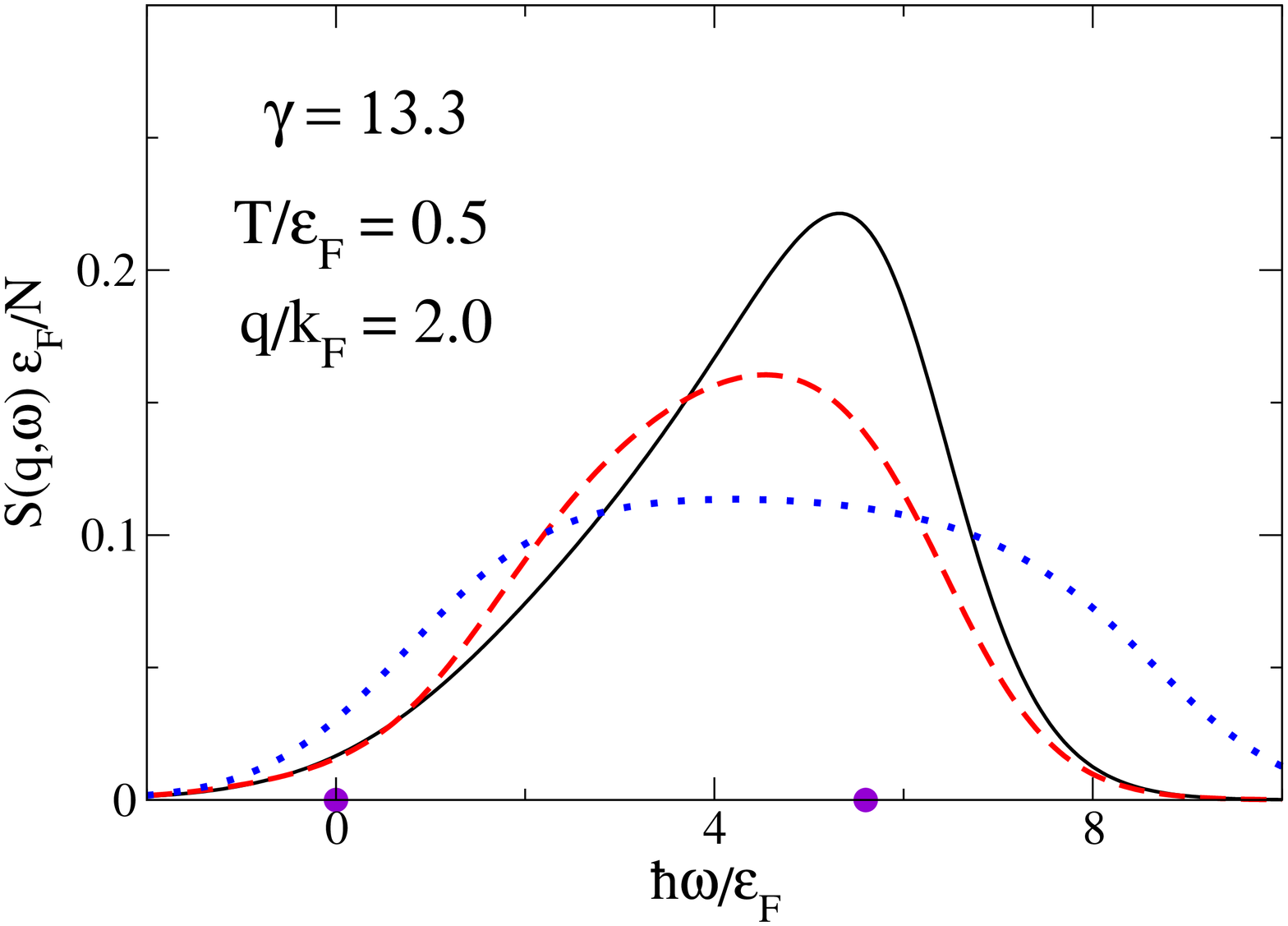}\\[7mm]
\includegraphics[width=.22\textwidth,angle=270,clip=on]{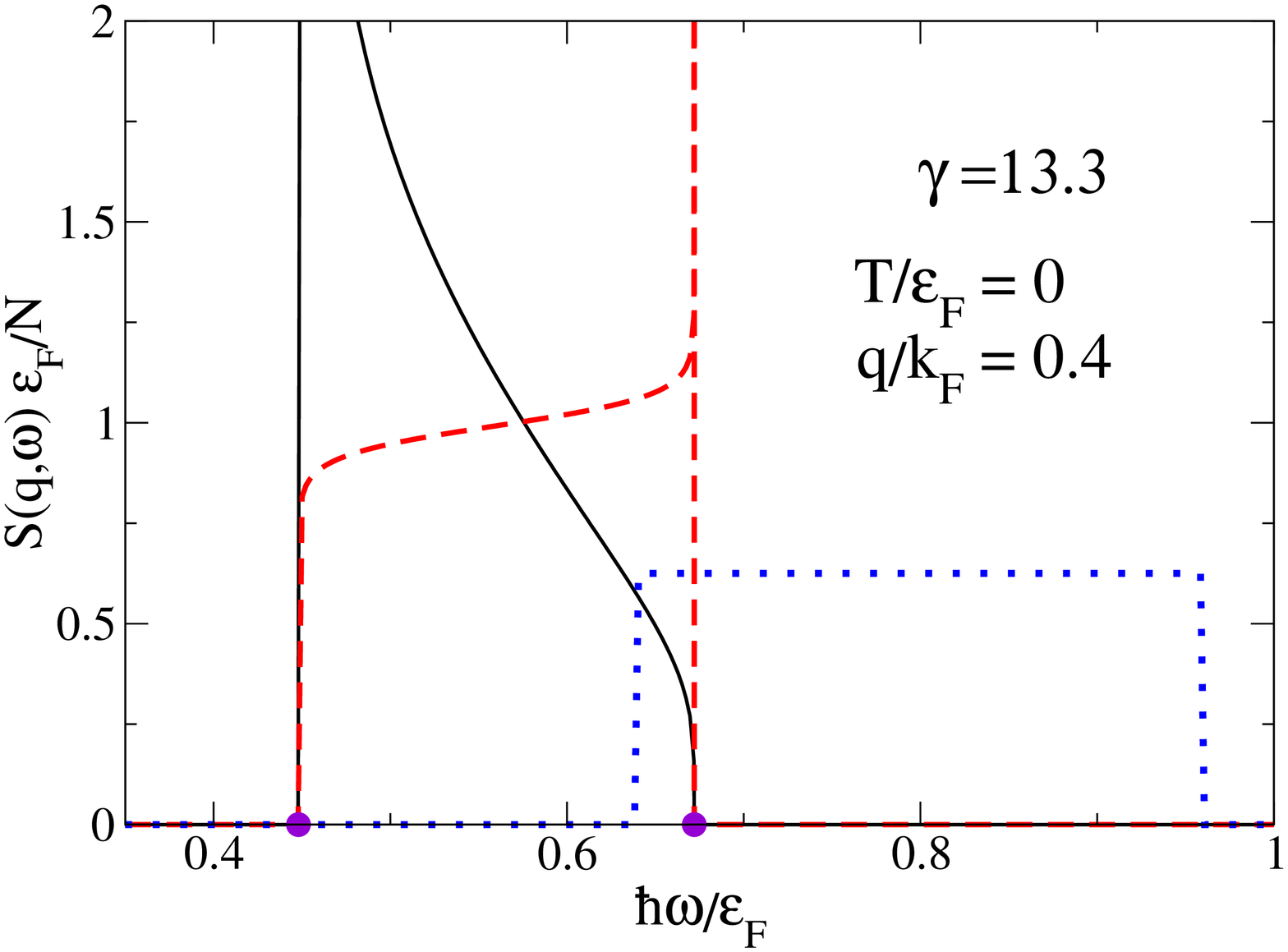}
\ \includegraphics[width=.22\textwidth,angle=270,clip=on]{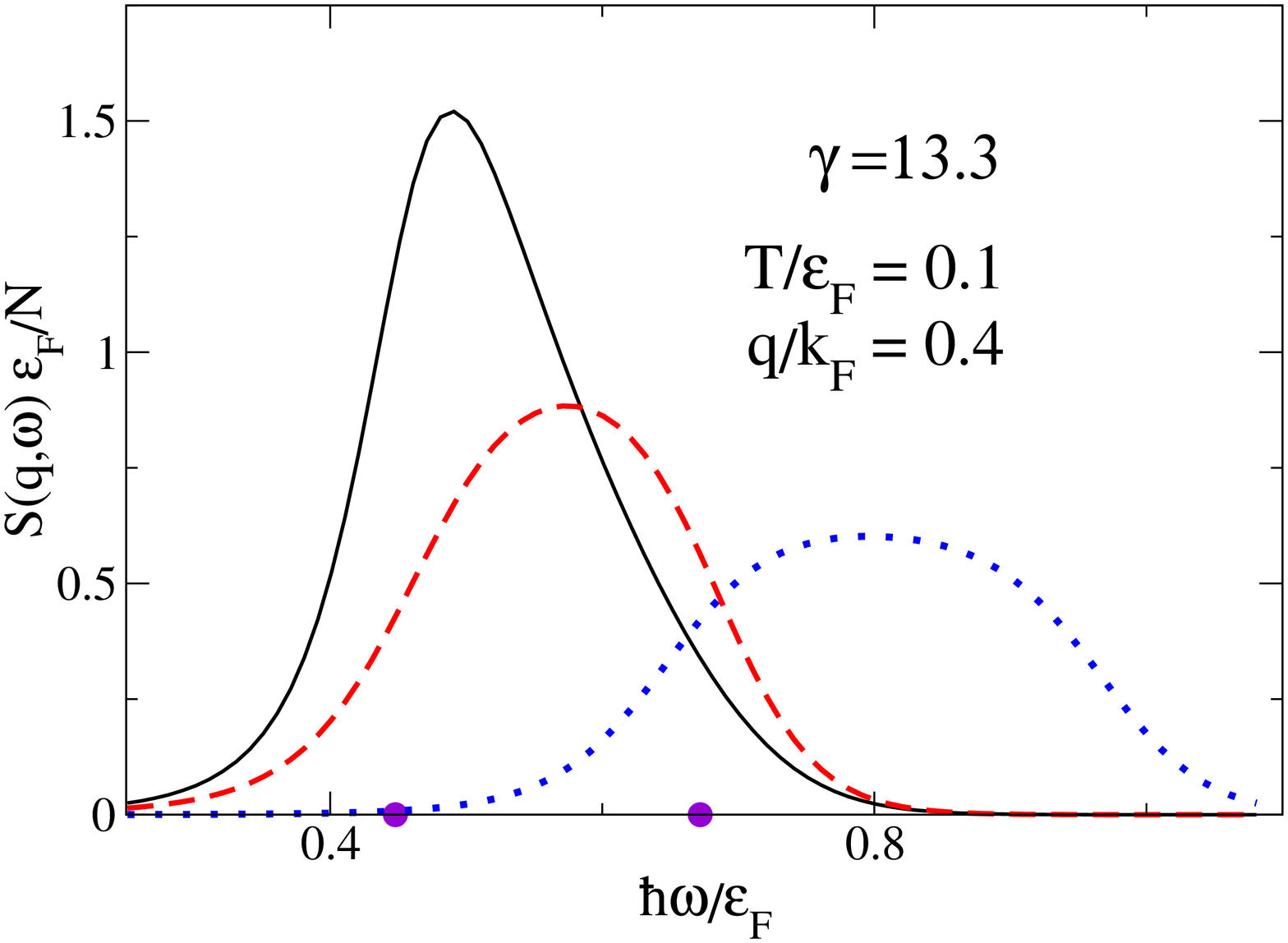}
\ \includegraphics[width=.22\textwidth,angle=270,clip=on]{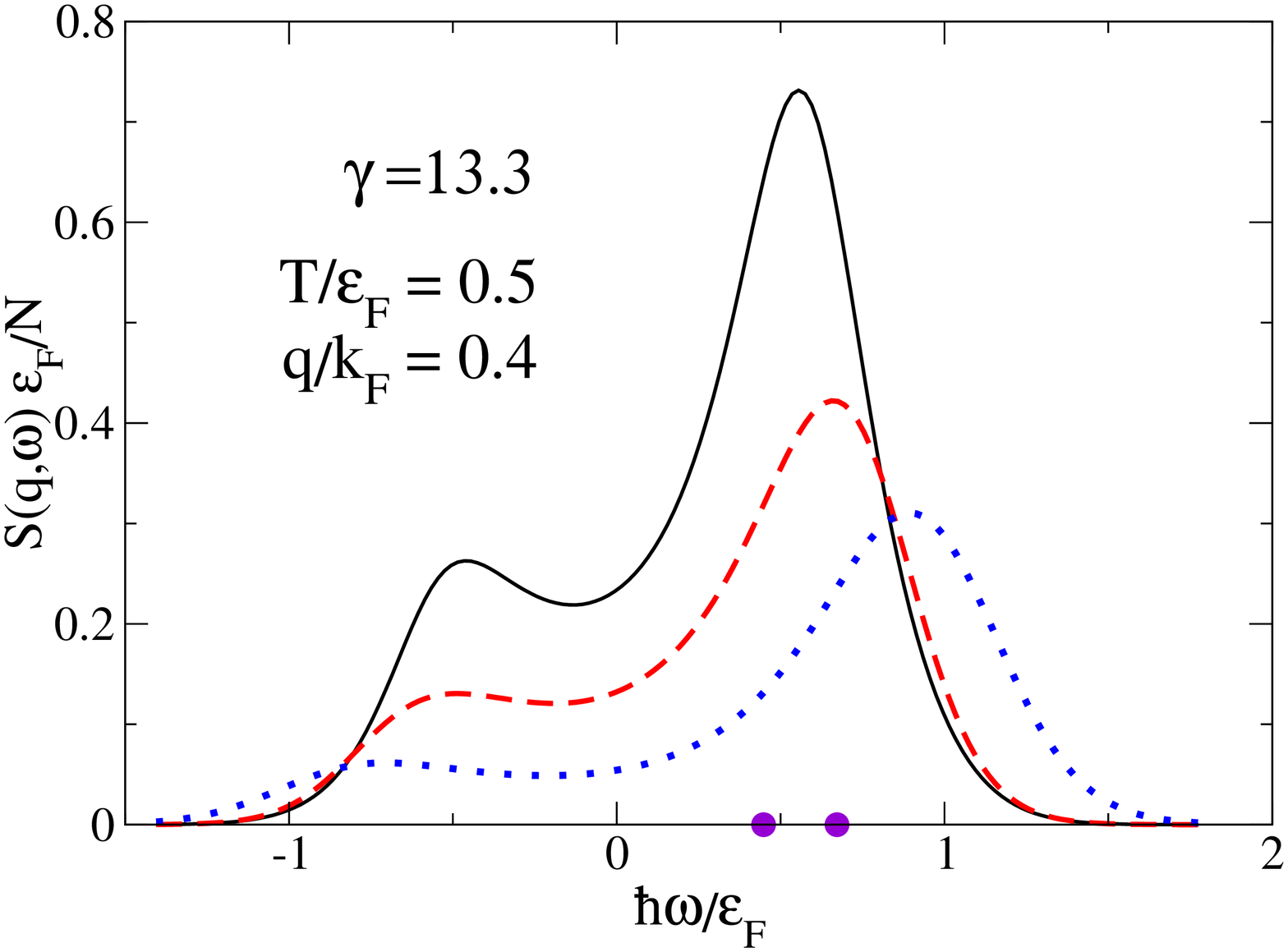}
\caption{\label{fig:DSFk2} (Color online) The DSF $S(q,\omega)\,\ef/N$ at
interaction strength $\gamma=13.33$ as a function of $\omega$ at $q=0.4 \kf$
and $q=2 \kf$ for various temperatures. The solid (black) line shows the RPA
result (\ref{eqn:strdyn}) or (\ref{eqn:strdyn_t}), the dashed (red) line shows
the first order expansion in $\gamma^{-1}$ (\ref{eqn:first-order}) or
(\ref{eqn:first-order_t}), and the dotted (blue) line shows the DSF of the TG
limit ($\gamma=\infty$) for comparison. The circular (violet) dots on the
x-axis denote $\omega_\pm$. The  first-order expansion always has a divergence
to $\pm\infty$ near $\omega_\pm$ at $T=0$, respectively, and the maximum is
shifted to $\omega_{+}$ at finite $T$. The RPA result, however, shows an
enhancement of low-energy excitation near $\omega_-$ at small $\q$. Note the
unphysical negative values of the DSF in the first order expansion near
$\omega_-$, in particular for the umklapp excitations at $q=2 \kf$ and $\omega$
close to zero.}
\end{figure*}

Due to a logarithmic singularity in $\chi_{1}^{(0)}$, the DSF vanishes on the
dispersion curves $\omega_\pm(q)$. For the TG gas at
$\gamma\to\infty$, the value of the DSF
within these limits is independent of $\omega$ and takes the value of $N m/(2
\pi \hslash^2 q n)$.
The energy-dependence in the RPA for finite $\gamma^{-1}$ is shown in
Figs.~\ref{fig:3D-DSF} and \ref{fig:DSFk2}. In particular, we see that
the umklapp excitations at $q=2 \kf$ and small $\omega$, which prohibit
superfluidity of the TG gas, are suppressed for finite $\gamma$. We
find that $S(2\kf,\omega)$ in the RPA approaches zero as
$1/\ln^2(\hslash\omega/\ef)$, in contrast to the
results of Refs.~\cite{castro_neto94,pitaevskii04}, which predict a power-law
dependence on $\omega$ for finite $\gamma$ based on a
pseudoparticle-operator approach.

In the RPA the enhancement of Bogoliubov-like excitations is seen as a strong
and narrow peak of the DSF in the RPA near $\omega_+$ at large momenta in
Fig.~\ref{fig:DSFk2}. At finite gamma and for small momenta $q\lesssim \pi n/2$,
however, the RPA
predicts a peak near $\omega_-$, in contrast to
the first-order result. Whether
this effect is real or an artefact of the RPA is not obvious and may be decided
by more accurate calculations or experiments. Spurious higher order terms in
the RPA and an improved approximation scheme have been discussed in
Ref.~\cite{brand98}. On the other hand, Roth and Burnett have recently observed
a qualitatively similar effect in numerical calculations of the DSF of the
Bose-Hubbard model \cite{roth04}.

The RPA result may be expanded in $1/\gamma$, which is
consistent with direct perturbation theory up to first order. This yields
for $\omega_{-}\leqslant \omega\leqslant \omega_{+}$
\be\label{eqn:first-order}
  S(q,\omega)\frac{\ef}{N} = \kf \frac{1+ 8\gamma^{-1}}{4 q} +
   \frac{\ln
  f(q,\omega)}{2\gamma} + {\cal O}(\gamma^{-2})
\ee
with $f(q,\omega) \equiv |(\omega^2 - \omega_-^2)/(\omega_+^2 - \omega^2)|$.
However, this first order expansion can assume negative values as seen in
Fig.~\ref{fig:DSFk2} although the DSF, given by Eq.~(\ref{sqomega}), should be
strictly non-negative, a property that  our RPA result (\ref{eqn:strdyn})
fulfills. Close to $\omega_+$, the first order expansion has a logarithmic
singularity tending to $+\infty$, which may be a precursor of the dominance of
Bogoliubov-like excitations in the DSF at small $\gamma$.
In the TG limit $\gamma^{-1}=0$, the DSF becomes discontinuous with
respect to $\omega$ at $\omega_{\pm}$
because the DSF of the TG gas is a step function [see
Eq.~(\ref{chi2})]. As a consequence, the first order
approximation~(\ref{eqn:first-order}) cannot be good for arbitrary values of
$q$ and $\omega$ but diverges in vicinity of $\omega_{\pm}$ due to  the slow
convergence of perturbation theory close to the point of discontinuity.
There is no formal problem here since the expression~(\ref{eqn:first-order}) remains
positive if for any given finite value of $q$  and
$\omega\not=\omega_{-}$ a large enough value of gamma is chosen.

Finally we discuss the $\delta$-function part of the DSF
(\ref{eqn:strdyn}). This contribution relating to discrete excitations
of collective character in the time-dependent HF scheme lies outside
of the continuum part and comes from possible zeros in the denominator
of $\chi(q,\omega+i\varepsilon)$. It is determined by the solution
$\omega_0(q)$ of the transcendental equation
$B=-D(q,\omega)\chi_{1}^{(0)}(q,\omega)$ in conjunction with
$\chi_{2}^{(0)}(q,\omega)=0$. We have solved this equation in various
limits and found that at most one solution for $\omega_0(q)$
exists. The strength $A(q)$ is given by the residue of the
polarizability at the pole $z_0=\omega_0(q)$. After small
algebra, we derive from Eq.~(\ref{chqom1}) \be
A(q)=N\frac{(\gamma-4)^{3}}{4(3\gamma-4)}\frac{\ef}{\hslash\omega_0(q)}\frac{\eta^{2}}{1+16\eta^{2}h_{0}},
\label{Aq}
\ee
where $\eta\equiv q/\kf$ and
\[
h_{0}\equiv\frac{\gamma(\gamma-6)^{2}}{3\gamma-4}
\frac{\big[\ln|\xi-(\eta-2)^{2}| -\ln|\xi-(\eta+2)^{2}|\big]^{-1}}
{[\xi-(\eta-2)^{2}][\xi-(\eta+2)^{2}]}
\]
with $\xi\equiv [\hslash\omega_0(q)/\ef]^{2}\gamma^{2}/[\eta(\gamma-4)]^{2}$.

\begin{figure}
\includegraphics[width=.8\columnwidth]{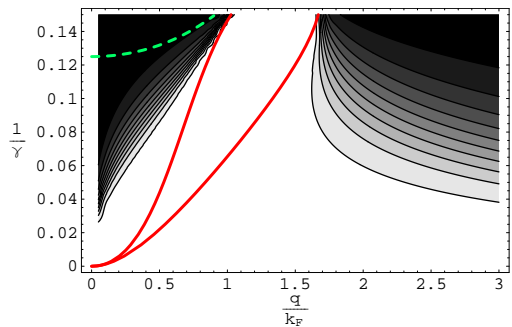}
\caption{\label{fig:SRCont} (Color online) Contour plot of $A(q)\, 2 m
\omega_0(q)/(\hslash N q^2)$, which shows the importance of the $\delta$-function
contribution. Values are given in ten contours between 0 (white) and 1 (black).
Between the thick (red) lines there is no discrete contribution because $\chi$
given by Eq.~(\ref{chqom1}) has no poles. Left of this region, there is a
discrete part with $\omega_0 < \omega_-$ and right of it there is one with
$\omega_0>\omega_+$. Above the dashed (green) line the RPA breaks down due to
an instability of the HF ground state as discussed in  Sec.~\ref{sec:hf}. }
\end{figure}

Numerical values for $A(q)\omega_0(q)$ are shown at finite $\gamma$
in Fig.~\ref{fig:SRCont}. For  small $q$ we find a $\delta$-function
contribution at
$\omega_0(q) < \omega_-$  whereas for large $q$ there is a discrete
contribution
at $\omega_0(q) > \omega_+$ (see
Fig.~\ref{fig:3D-DSF}a). In the limit
$\q \to \infty$ at finite $\gamma$, the $\delta$-part completely determines the
DSF as the continuum part vanishes; asymptotically  $A\simeq N$, and
$\omega_0 \simeq \hslash q^2/(2 m)$ becomes  the free particle dispersion,
reminiscent of the DSF for the weakly interacting Bose gas at large momentum
in Bogoliubov theory \cite{bog47}.

For small $\gamma^{-1}$, the strength $A(q) \simeq 2 N \gamma
\exp(-\gamma\q/\kf)$ is exponentially suppressed and possible
solutions are close to the dispersion branches $\omega_\pm$ with
$|\omega_0-\omega_\pm| \propto \exp(-\gamma\q/\kf)$. Due to this
proximity of the discrete and continuous parts and expected smearing
of discrete contributions by interactions beyond the RPA, we may
conjecture that the $\delta$-function should be seen as part of the
continuum, enhancing contributions near the border.  Moreover, at
finite temperatures there is no $\delta$-function contribution even
within the RPA, as we discuss in Sec.~\ref{sec:DSF_T} below. Indeed, we
know from the exact solutions~\cite{lieb63:2} that the energy spectrum is
continuous.

The RPA polarizability
(\ref{chqom1}) is a retarded Green's function and thus has to be
analytic in the upper half complex plane
\cite{pines89,pines61}. At zero temperature, the analyticity breaks
down above the dashed (green) line in Fig.~\ref{fig:SRCont}. The
instability of the RPA results from the instability of the HF
approximation~\cite{thouless61} and arises exactly at the critical
value of $\gamma$ when the isothermal speed of sound equals to zero,
see Fig.~\ref{fig:vsound}.

\subsubsection{Finite temperatures}
\label{sec:DSF_T}

At finite temperatures we obtain the DSF by means of Eqs.~(\ref{chqom1})
and (\ref{imchisqomega})
\be
  S(q,\omega)=\frac{ \chi^{(0)}_{2}(q,\omega) B [1-\exp(-\beta\hslash\omega)]^{-1}} {\pi (1-4\,\gamma^{-1})
               \left[\Big(B+D\chi_{1}^{(0)}\Big)^2 +
               \Big(D\chi_{2}^{(0)}\Big)^2\right]},
\label{eqn:strdyn_t}
\ee
where the real and imaginary parts of the polarizability $\chio_{1,2}$
are given by Eqs.~(\ref{chi1t}) and (\ref{chi2t}), respectively. The DSF is
shown in Figs.~\ref{fig:3D-DSF} and \ref{fig:DSFk2}. The main effect
of finite temperature is a smoothing of the zero-temperature features.
The $\delta$-function contribution to the DSF
disappears, since $\chi_{2}^{(0)}\not=0$ and thus the denominator of
Eq.~(\ref{eqn:strdyn_t}) does not vanish for $\omega\not=0$. It is
absorbed by the
continuum part of the DSF.

At finite temperatures, the non-vanishing contributions of the DSF
spread considerably beyond the particle-hole excitation spectrum
limited by $\omega_{-}$ and $\omega_{+}$ because the DSF no longer
probes the ground state but a thermal ensemble [see Eq.~(\ref{sqomega})].
For
negative values of the frequency, the DSF decays exponentially in
accordance with Eq.~(\ref{imchisqomega}). Similar to the case of zero
temperature, the enhancement of excitations still takes place close to
$\omega_{+}$ for $q\gtrsim \pi n/2$ and $\omega_{-}$ for $q\lesssim
\pi n/2$ at small values of temperature $T\lesssim 0.5\ef$.

To the first order in $\gamma^{-1}$, we have
\begin{align}
\label{eqn:first-order_t}
  S(q,\omega)\frac{\ef}{N} =&\ \frac{n_{q_{-}}-n_{q_{+}}}{1-\exp(-\beta\hslash\omega)}
  \bigg[\kf \frac{1+ 8\gamma^{-1}}{4q} \nonumber\\
   &\ +   \frac{1}{2\gamma}{\rm P}\int\d k\,\frac{n_{k+q_{+}}-n_{k+q_{-}}}{k}\bigg]
   + {\cal O}(\gamma^{-2}).
\end{align}
This linear approximation fails in vicinity of the umklapp excitation
$q=2\kf$ and $\omega=0$, yielding unphysically negative values of the DSF
contrary to the obtained RPA expression~(\ref{eqn:strdyn_t}), see
Fig.~\ref{fig:DSFk2}.

\subsubsection{Sum rules for the DSF}
\label{sec:sum_rules}

Sum rules for the DSF are an important test for checking the validity of the
obtained expressions.
In particular, the $f$-sum rule~\cite{pitaevskii03:book,pines61,pines89}
\be
m_1\equiv \hslash^2 \int
\omega  S(q,\omega) \d \omega = N \hslash^2 q^2/(2 m)
\label{fsumrule}
\ee should be fulfilled to all orders in $\gamma^{-1}$ within the
RPA~\cite{thouless61}. We have verified it by numerical integration
and found excellent agreement at finite values of $\gamma$ for both
the zero-temperature DSF (\ref{eqn:strdyn}) and the finite temperature
expression (\ref{eqn:strdyn_t}).  The $f$-sum rule can also be
verified analytically from the large $\omega$
asymptotics $\chi(q,\omega)\simeq -2m_{1}/(\hslash \omega)^2$
using  Eqs.~(\ref{chqom1}) and (\ref{chiqomandsqom}), assuming
that $\chi$ is analytic as a function of $\omega$ in the upper
half complex plane.

The sum rule for the isothermal compressibility~\cite{pitaevskii03:book}
\be
\lim_{q\to 0} {\rm P}\int
\frac{S(q,\omega)}{\omega} \d \omega =
\frac{N}{2n}\Big(\frac{\partial n}{\partial \mu}\Big)_{T}
\label{sumrule1}
\ee
holds also to all orders
in $\gamma^{-1}$, which can be checked
analytically. Indeed, by comparing Eq.~(\ref{chqom1}) with the HF isothermal
compressibility discussed in Sec.~\ref{sec:hf} we derive
\be
\lim_{q\to 0}\chi(q,0+i\varepsilon)
=\frac{N}{n}\Big(\frac{\partial n}{\partial \mu}\Big)_{T}.
\label{compress}
\ee
Then Eq.~(\ref{sumrule1}) is a direct consequence of the dispersion relation
(\ref{chiqomandsqom}) at $\omega=0$.

\subsubsection{Consequences for superfluidity}

As we have argued in Ref.~\cite{brand05}, the value of the DSF near the
umklapp excitations at $\omega =0$ and $q=2\kf$ is relevant for the
phenomenon of superfluidity according to the Landau criterion. A finite
value of $S(q,\omega)$ will prohibit persistent currents, as
spontaneous excitations initiated by infinitesimal perturbations would
be able to dissipate the translational kinetic energy stored in the
current. Our finite temperature results clearly show that
$S(q=2\kf,\omega=0)$ at a given value of $T>0$ will always be positive
and finite for large enough $\gamma$ in both the RPA expression
(\ref{eqn:strdyn_t}) and the first order expansion
(\ref{eqn:first-order_t}). We thus conclude that there is no
superfluidity in the large-$\gamma$ regime at finite temperatures, in
accordance with Popov's analysis \cite{popov72} made many years ago.
The question of efficient suppression of the DSF in the vicinity of the
umklapp excitation at $T=0$ cannot be fully answered within the present
approach due to the nonregularity of $S(q,\omega)$ at this point, as
discussed in Ref.~\cite{brand05}, and will be left to future
investigations.

\subsection{Static structure factor and pair distribution function}
\label{sec:ssf}

The static structure factor
$S(q)$~\cite{pitaevskii03:book,pines61,pines89}  is a function of
momentum only and is obtained by integrating the DSF over the
frequency
\be
S(q)\equiv\frac{\langle\hat{\rho}_{q}\hat{\rho}^{\dag}_{q}\rangle}{N}=
\frac{\hslash}{N}\int \d\omega\,S(q,\omega).
\label{strstat}
\ee
The results of numerical integration of the DSF in the RPA are plotted
in Fig.~\ref{fig:SSF}.

The static structure factor contains information about the static
correlation properties of a
system and directly relates to the pair distribution function or the
normalized density-density correlator
$g(x)
\equiv\langle{\hat\Psi}^\dag(x){\hat\Psi}^{\dag}(0){\hat\Psi}(0){\hat\Psi}(x)\rangle
/ n^{2}$
by the equation
\begin{equation}
g(x)=1+\int \frac{\d q}{2\pi n}\,e^{iqx}\big[S(q)-1\big].
\label{sgx}
\end{equation}

At small momenta $S(q)$ can be related to the isothermal compressibility
because the main contribution into the integral (\ref{strstat}) comes from the
``classical'' region $\hslash\beta\omega\ll 1$~\cite{pitaevskii03:book}.
We derive from Eqs.~(\ref{chiqomandsqom}), (\ref{imchisqomega}),
(\ref{compress}), and (\ref{strstat}) \be
\lim_{q\to0}S(q)=\frac{T}{n}\Big(\frac{\partial n}{\partial
\mu}\Big)_{T}=\frac{T}{m v^{2}_{\rm T}}.
\label{strstatzero}
\ee
This relation of the structure factor $S(q)$ to the speed of sound
$v_{\rm T}$ at small momentum implies that $S(q)$ in the RPA is exact up
to first order in $\gamma^{-1}$ and is overestimated at finite
$\gamma$ as seen from the results for $v_{\rm T}$ in Fig.~\ref{fig:vsound}.

\subsubsection{Zero temperature}

We can obtain the static structure factor
(\ref{strstat}) from Eq.~(\ref{eqn:first-order}) to the first order
\begin{equation}
S(q)=S^{(0)}(q)+\gamma^{-1} S_{1}(q) + \mathcal{O}(\gamma^{-2}).
\label{strstatfirst}
\end{equation}
Here $S^{(0)}$ denotes the static structure factor for the ideal 1D Fermi gas
\begin{equation}
S^{(0)}(q)=\left\{ \begin{array}{ll}
|q|/(2\pf),&  |q| \leqslant 2\pf,\\[2mm]
1,     & |q| \geqslant 2\pf,
\end{array}\right.
\label{strstat0}
\end{equation}
and the function $S_{1}(q)$ takes the form
\begin{align}
S_{1}(q)=\ &|\eta|\big[ r(\eta)- |\eta-2|\ln|\eta-2| -|\eta+2|\ln |\eta+2| \big]\nonumber\\
         &\ +4S^{(0)}(q)
\label{strstatcorr}
\end{align}
with the dimensionless wave vector $\eta\equiv q/\pf$ and the function
\begin{equation}
r(\eta)\equiv\left\{ \begin{array}{ll}
4\ln 2,               & |\eta|\leqslant 2,\\[2mm]
2|\eta|\ln|\eta|,     & |\eta|\geqslant 2.
\end{array}\right.
\label{reta}
\end{equation}
\begin{figure*}
\includegraphics[width=.22\textwidth,angle=270,clip=on]{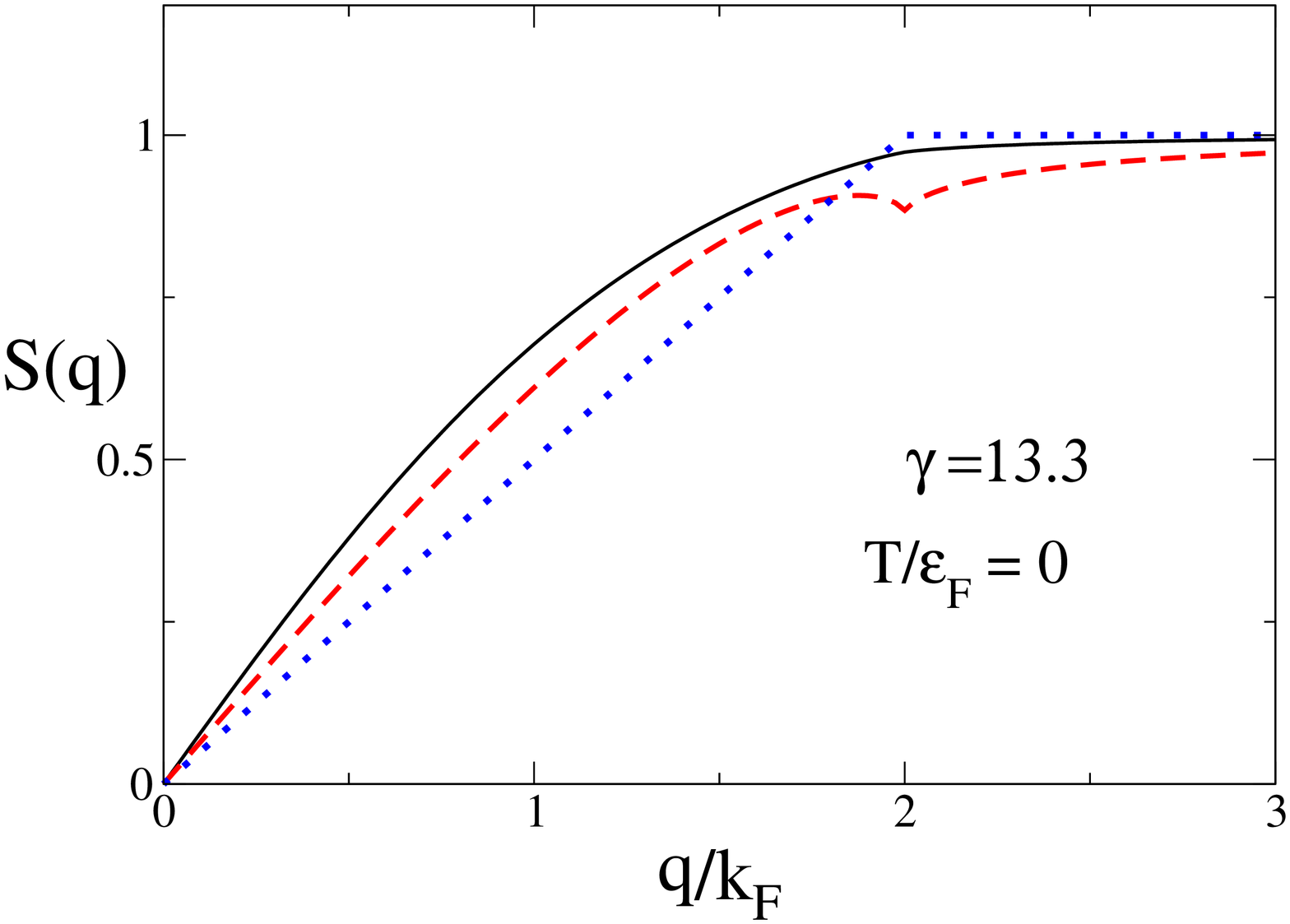}\
\includegraphics[width=.22\textwidth,angle=270,clip=on]{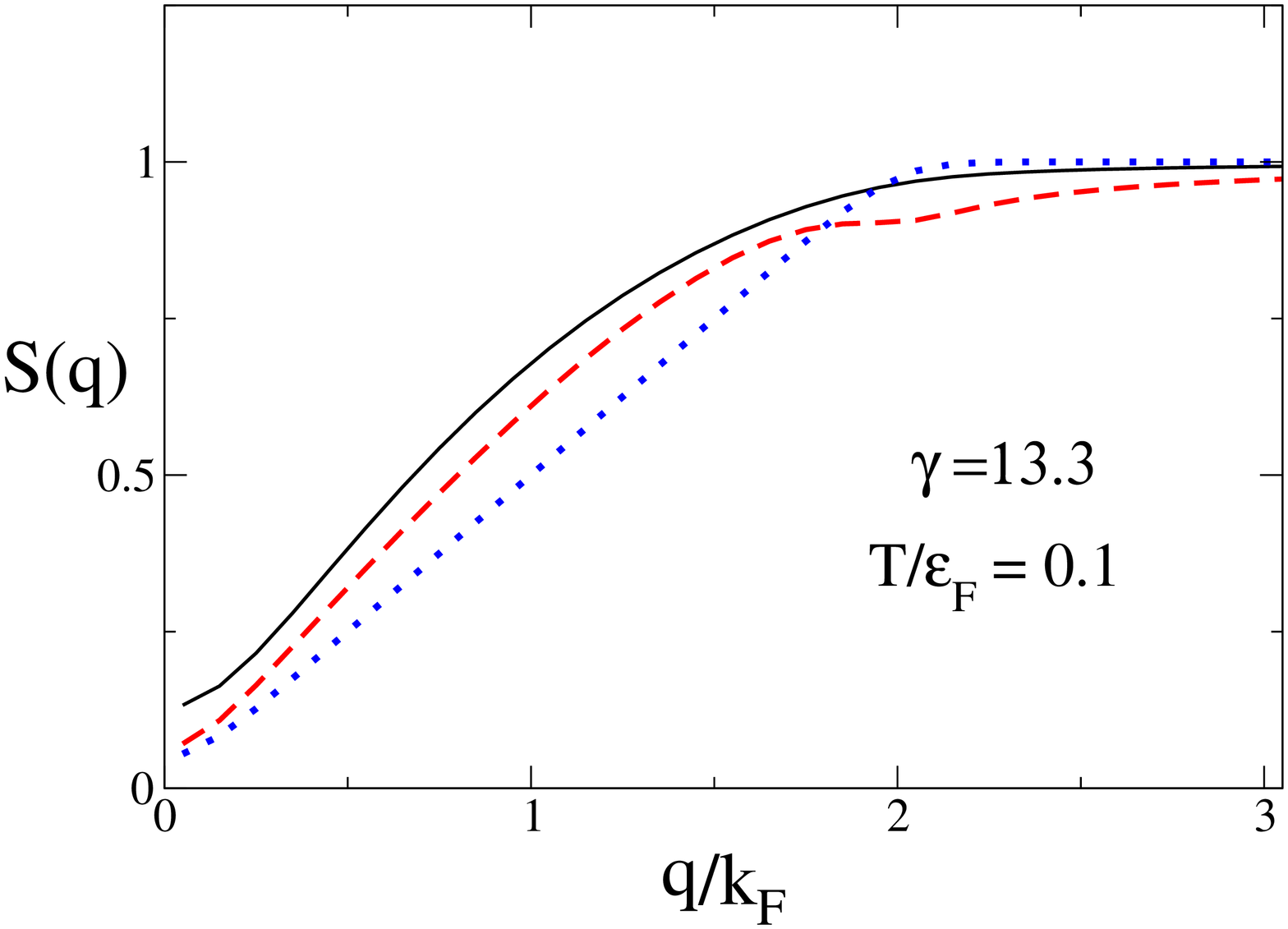}\
\includegraphics[width=.22\textwidth,angle=270,clip=on]{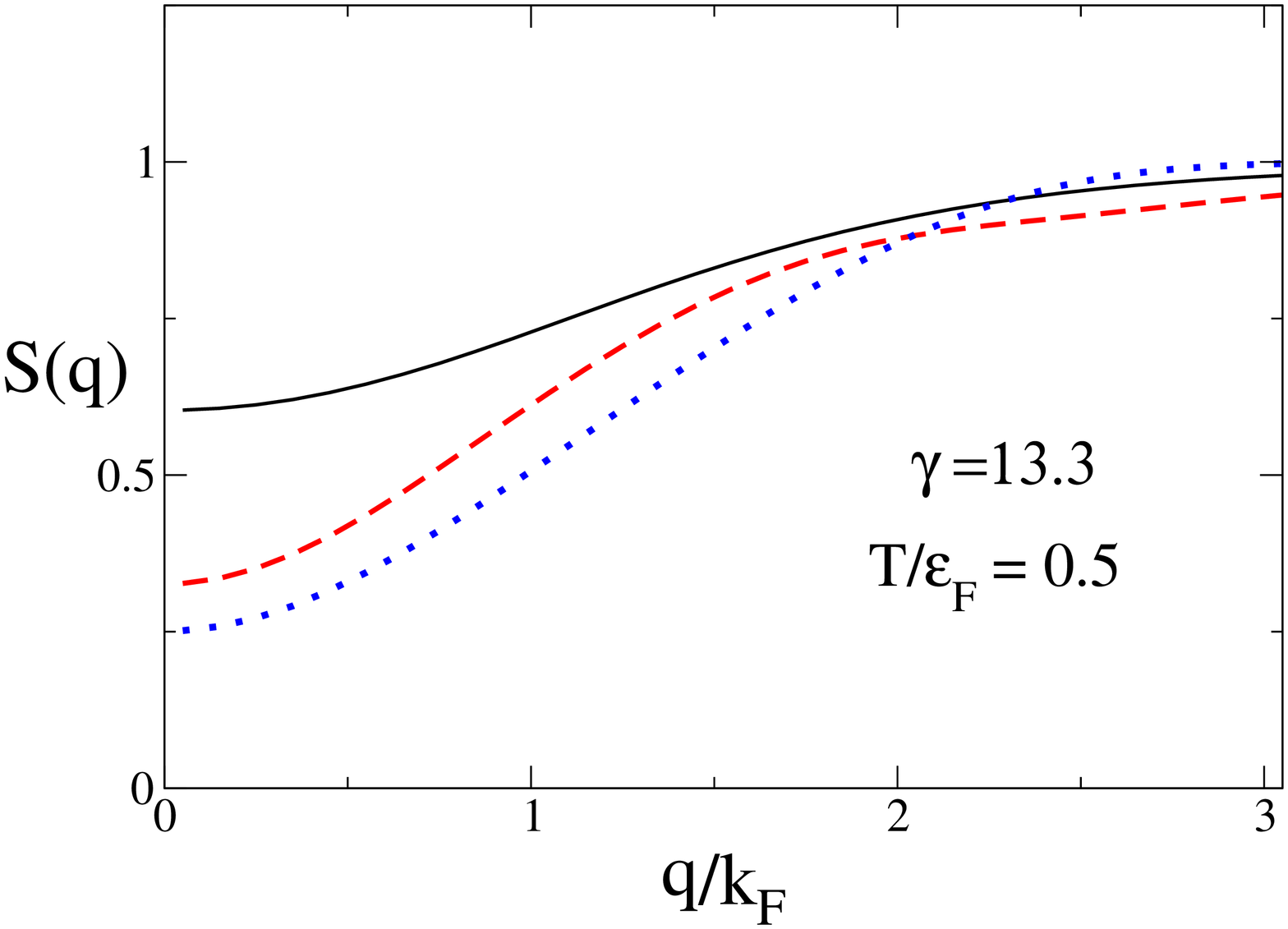}
\caption{\label{fig:SSF} (Color online) The static structure factor $S(q)$ as a
function of momentum for $\gamma=13.3$ and various temperatures. The solid
(black) line shows the RPA result obtained by numerical integration from
Eqs.~(\ref{eqn:strdyn}) or (\ref{eqn:strdyn_t}). The dashed (red) line shows
the first order expansion (\ref{strstatfirst}) in $\gamma^{-1}$ with
Eqs.~(\ref{strstat0}) and (\ref{strstatcorr}) at zero temperature or
Eqs.~(\ref{strstat0_t}) and (\ref{strstatcorr_t}) at non-zero temperatures. The
dotted (blue) line shows the static structure factor in the TG limit
($\gamma=\infty$) (\ref{strstat0}) or (\ref{strstat0_t}) for comparison. Note
the unphysical cusp at $q=2\kf$ in the first-order expansion of the static
structure factor at zero temperature and some traces of it at small
temperature.}
\end{figure*}

The obtained correction $S_{1}(q)$ is continuous and has the
asymptotics $S_{1}(q)\simeq-8/(3\eta^2)+\mathcal{O}(1/\eta^4)$ when
$\eta\to\infty$ and $S_{1}(q)\simeq2|\eta|
-|\eta|^3/2+\mathcal{O}(\eta^5)$ when $\eta\to0$. The latter
asymptotics gives us a possibility to determine the coupling constant
$\gamma$ experimentally from the phonon part of the static structure
factor (\ref{strstatfirst}) for $q\lesssim\pf$
\begin{equation}
S(q)\simeq\frac{|q|}{2\pf}\bigg(1+\frac{4}{\gamma} -\frac{q^2}{\gamma\pf^2}\bigg).
\label{strstatfirst_exp}
\end{equation}
The static structure factor at small $q$ is
related to the sound velocity~\cite{pitaevskii03:book} by
$S(q)\simeq{\hslash |q|}/{(2 m v_{\rm T})}$. It is easily seen that our
result (\ref{strstatfirst_exp}) is consistent with the sound velocity of
Eq.~(\ref{eqn:speedLL}).

Figure \ref{fig:SSF} shows the static structure factor in the full RPA
and its first order expansion (\ref{strstatfirst}) at $\gamma=13.3$
and the TG
limit $S^{(0)}$ for comparison. The first order result shows
a cusp which is an artefact of the first order expansion.

Using Eq.~(\ref{sgx}) in conjunction with relations
(\ref{strstatfirst})-(\ref{reta}), we can represent our result for
the pair distribution function in the form
\begin{align}
g&(x)=\ 1-\frac{\sin^{2}z}{z^{2}}-\frac{2\pi}{\gamma}\frac{\partial}{\partial z}\frac{\sin^{2}z}{z^{2}}
-\frac{4}{\gamma}\frac{\sin^{2}z}{z^{2}}\nonumber\\
& +\frac{2}{\gamma}\frac{\partial}{\partial z}
\left[\frac{\sin z}{z}\int_{-1}^{1}\d\eta\,\sin(\eta z)
\ln\frac{1+\eta}{1-\eta}\right] + \mathcal{O}(\gamma^{-2}),
\label{gxfirstour}
\end{align}
where $z=\kf x=\pi n x$. It follows from this equation that $g(x=0)$ vanishes
not only in the TG limit but also in first order of $\gamma^{-1}$, which is
consistent with the results of Refs.~\cite{lieb63:1,Gangardt2003a}
and the HF expression (\ref{gx0hf}) below, indicating once more the validity
of our results.
The physically correct limit $g(x)\to 1$ for $x\to\infty$ is fulfilled due to
Eq.~(\ref{sgx}). A
similar expression for $g(x)$ was derived in Ref.~\cite{korepin93} for
the large distance asymptotics. To our knowledge,
Eq.~(\ref{gxfirstour}) shows for the first time the full $x$
dependence of $g(x)$ up to first order in $\gamma^{-1}$.

\subsubsection{Finite temperatures}

The first-order approximation for the static structure factor at finite
temperatures is obtained by using Eqs.~(\ref{eqn:first-order_t}) and
(\ref{strstat}).
The result takes the form of Eq.~(\ref{strstatfirst})
but with the TG, or free Fermi, static structure factor
\begin{equation}
S^{(0)}(q)=\frac{m}{2\hslash\kf q}\int\d \omega\frac{n_{q_{-}}-n_{q_{+}}}{1-\exp(-\beta\hslash\omega)}
\label{strstat0_t}
\end{equation}
and with the function $S_{1}(q)$
\begin{align}
S_{1}(q)\!=&\frac{m}{\hslash\kf^2}\int\d \omega \frac{n_{q_{-}}-n_{q_{+}}}{1-\exp(-\beta\hslash\omega)} {\rm P}\!\int\d
k\,\frac{n_{k+q_{+}}-n_{k+q_{-}}}{k} \nonumber\\
&+4S^{(0)}(q),
\label{strstatcorr_t}
\end{align}
where $q_{\pm}$ is given by Eq.~(\ref{kpm}) at $\gamma^{-1}=0$.  The
finite temperature results are plotted in Fig.~\ref{fig:SSF}. One can
see an unphysical behaviour of the first order approximation near
$q=2\kf$, in contrast to the full RPA result.
The small momentum limits for  $S(q)$ are determined by the speed of
sound $v_{\rm T}$ through Eq.~(\ref{strstatzero}).
In the plot on the
right hand side, the RPA overestimates $S(q)$ at small $q$ as a result
of the deviations of the speed of sound as seen in Fig.~\ref{fig:vsound}.

\subsubsection{Limits of validity}

When the interaction is proportional to a small parameter, the RPA
method is applicable and yields correct values of the DSF at least up
to the first order in this parameter
\cite{pines89,pines61,thouless72}. This implies the validity of the
obtained expressions for the polarizability, the dynamic and static
structure factors, and the pair distribution function up to the first
order in $\gamma^{-1}$. Smallness of the inverse coupling constant
means, in particular, small values of the pair distribution
function in the contact point: $g(x=0)\ll 1$, which is the TG regime
by definition.

A classification of different regimes in the 1D Bose gas for
arbitrary temperatures was given in Ref.~\cite{Shlyapnikov2003}. As
it was mentioned above, it is possible to obtain the values of $g(x)$
at $x=0$ from the exact solution of the Lieb-Liniger model with the
help of the Hellmann-Feynman theorem. The TG regime is
realized~\cite{Shlyapnikov2003} when
\be
\gamma \gg \text{max}(1,\sqrt{T/\ef}),
\ee
which gives also the criterion of validity of the RPA results.
We can derive this criterion  within the HF approach
of Sec.~\ref{sec:hf}. Indeed, we obtain the following result for the
pair distribution function by applying the
Hellmann-Feynman theorem to the HF grand potential $\Omega$:
\be
g(x=0)=\frac{4\pi^2}{\gamma^2}\frac{\langle k^2\rangle}{\kf^2}.
\label{gx0hf}
\ee
Using the low-temperature expansion of the average
momentum (\ref{q2av}) $\langle k^2\rangle =(\kf^2/{3}) [1
+\pi^2T^2/(4\ef^2) +\cdots]$ and the high-temperature expansion $\langle
k^2\rangle =\kf^2 T/[2\ef(1-4\gamma^{-1})] +\cdots$, we arrive at the
above mentioned restriction on $\gamma$.

The validity of the RPA requires, in particular, the stability of the
HF solutions \cite{thouless61}.
Thus our results are applicable in
practice for $\gamma\gtrsim 10$, see discussion in Sec.~\ref{sec:hf}.

Note that the HF expression (\ref{gx0hf}) yields the correct value of the pair
distribution function only at $x=0$ but up to the second order in
$\gamma^{-1}$. This is due to validity of the HF approximation in the first
order in $1/\gamma$; hence, the derivative with respect to $\gamma$ of the
first-order correction for the grand potential gives the correct value of
$g(x=0)$, proportional to $1/\gamma^2$. By contrast, the RPA expression
(\ref{gxfirstour}) and its finite temperature generalization yield the values
of $g(x)$ for arbitrary $x$ but guarantee validity only up to the first order.
For this reason, the numerical values of $g(x=0)$ obtained with
Eqs.~(\ref{eqn:strdyn_t}), (\ref{strstat}), and (\ref{sgx}) differ from those
of Eq.~(\ref{gx0hf}).

\section{Conclusion}
\label{sec:sum}

We have derived variational approximations for the
dynamic polarizability and related two-particle correlation functions of the
one-dimensional Bose gas, extending our previous results
\cite{brand05} to finite temperatures. The approximations are good for
strong interactions and yield expansions valid to first oder in
$1/\gamma$, which had not been available previously. We have carefully
checked the consistency with known
limits and sum rules and analyzed the limits of validity of the
derived equations.
Due to the Bose-Fermi duality, our results are
equally applicable for strongly interacting bosons as well as for
weakly-interacting spinless fermions.
Our result for the DSF indicates a dramatic departure from the TG
limit already for very small values of $1/\gamma$ by enhancing
Bogoliubov-like excitations and by suppressing umklapp excitation, which
are the main obstacle to observing superfluid-like response in the 1D
Bose gas. However, we find that superfluidity at finite temperatures
is strictly prohibited in the large-$\gamma$ regime as umklapp
excitations are always associated with a finite probability. Finite
temperature effects generally are found to smear out
the sharp features of the zero temperature correlation
functions. Nevertheless, at a level of 10\% of the Fermi temperature
$\varepsilon_{\rm F}$, the main effects should be well observable in
experiments.

Our results also establish the usefulness and validity of the
fermionic pseudopotentials (\ref{eqn:SenPP}) and (\ref{eqn:our}) and
the variational Hartree-Fock approximation and RPA. The method can
easily be extended to further studies in the large-$\gamma$ regime by
including the effects of harmonic or periodic external potentials or
by studying nonlinear response properties. Furthermore, the acquired
knowledge of the dynamic density correlations will be useful for
constructing an accurate time-dependent density functional theory,
extending the approach of Ref.~\cite{brand04a}.

The authors are grateful to Sungyun Kim and Rashid Nazmitdinov for useful remarks.

\bibliographystyle{prsty}

\begin{thebibliography}{99}
\bibitem{note_bragg} One can argue (see Ch.~12 of
Ref.~\cite{pitaevskii03:book}) that the Bragg scattering gives us the imaginary
part of the polarizability rather than the DSF itself. This can be essential
at finite temperature.  Certainly, knowing one quantity, one can easily
calculate the other by relation~(\ref{imchisqomega}).
\bibitem{note_local} Note that an arbitrary local interaction $V(x_1-x_2)$ is
associated with the kernel $V(x_1,x_2;x'_2,x'_1) =V(x_1-x_2)
\delta(x_1-x'_1)\delta(x_2-x'_2)$.
\end{thebibliography}

\end{document}